\documentclass[conf]{new-aiaa}
\usepackage{graphicx}      % include this line if your document contains figures
\usepackage{natbib}        % required for bibliography
\usepackage{xcolor}

\usepackage{graphicx}
\usepackage{amsmath}
\usepackage[version=4]{mhchem}
\usepackage{longtable,tabularx}
\usepackage{bm}
\usepackage{float}
\usepackage{algorithm}
\usepackage{algpseudocode}
\usepackage{multirow}
\usepackage{subcaption}
\usepackage{textcase}

\title{Quadrotor Takeoff Trajectory Planning in a One-Dimensional Uncertain Wind-field Aided by Wind-Sensing Infrastructure}
\author{Nicholas Kakavitsas\footnote{Graduate Student, Department of Mechanical Engineering and Engineering Science, nkakavit@charlotte.edu,  AIAA Student Member.}, Artur Wolek\footnote{Assistant Professor, Department of Mechanical Engineering and Engineering Science, awolek@charlotte.edu, AIAA Member.}}
\affil{
University of North Carolina at Charlotte, Charlotte, North Carolina, 28223}

\begin{document}
%\IEEEoverridecommandlockouts
% \overrideIEEEmargins deprecated version 1.7+
\maketitle

\begin{abstract}
This paper investigates optimal takeoff trajectory planning for a quadrotor modeled  with  vertical-plane rigid body dynamics in an uncertain, one-dimensional wind-field. The wind-field varies horizontally and  propagates across an operating region with a known fixed speed. The operating area of the quadrotor is equipped with wind-sensing infrastructure that shares noisy anemometer measurements with a centralized trajectory planner. The measurements are assimilated via Gaussian process regression to predict the wind at unsampled locations and future time instants. A minimum-time optimal control problem is formulated for the quadrotor to take off and reach a desired vertical-plane position in the presence of the predicted wind-field. The problem is solved using numerical optimal control. Several examples illustrate and compare the performance of the trajectory planner under varying wind conditions and sensing characteristics.
%assuming the estimated wind-field and simulated in the actual wind-field.
%The benefits of providing a wind-estimate to a numerical optimal control solver, GPOPS-II, are evaluated via simulation trials in wind-fields with varying strengths and variances, along with estimates generated with both high and low quality measurements.  To evaluate the trials, the control generated by the solver is re-simulated using the same wind information that was provided to the solver, to confirm the result is physically valid.  After confirming the result, the calculated control is used to propagate the dynamics through the true wind-field.  The final state provided by the trajectory generated from the optimal control solver and a separate ODE45 solver are compared via calculating the euclidean distance error.  As expected, in a strong and highly varying wind-field, with low quality measurements, the error was high, but in all other tested cases, the error was low enough to be a valid starting point for real-world experimentation.
\end{abstract}

\section{Nomenclature}

{\renewcommand\arraystretch{1.0}
\noindent\begin{longtable*}{@{}l @{\quad=\quad} l@{}}
$O$ & origin of inertial reference frame \\
$\mathcal{I}$& inertial reference frame; $\{ O, {\bm i}_1, {\bm i}_2, {\bm i}_3 \}$ \\
%
%$\bm{i}_{1-3}$ & Orthonormal unit vectors oriented in the north-east-down directions of the inertial reference frame, respectively. \\
$G$ & center of mass of the quadrotor \\
$\mathcal{B}$& body reference frame; $\{ G, {\bm b}_1, {\bm b}_2, {\bm b}_3 \}$ \\
%$\bm{b}_{1-3}$ & Orthonormal unit vectors oriented in the forward-right-down body directions, respectively. \\
$l$ & distance between propellers \\
%\bm{W}$ & Weight vector \\
%$\bm{D}$ & Drag vector \\
$T_{\rm r}, T_{\rm f}$ & equivalent thrust for two rear and forward motors, respectively \\
$T_{\rm{max}}$ & maximum thrust limit\\
%$\left[\cdot\right]^{\text T}$ & Transpose operation\\
$\bm{v}$ & inertial velocity \\
$\bm{v}_{{\rm r}}$ & flow-relative velocity \\
$\bm{v}_{\delta}$ & wind velocity \\
$\bm{\Theta}$ & vector of roll, pitch, and yaw Euler angles, respectively; $\left[\phi,\theta,\psi\right]^{\rm T}$  \\
$\bm{R}(\bm{\Theta})$ & yaw-pitch-roll rotation matrix \\
%$c(\cdot)$ & Short form of $\cos$ \\
%$s(\cdot)$ & Short form of $\sin$ \\
%$SO(3)$ & Special orthogonal group \\
$\bm{x}$ & inertial position vector of the quadrotor expressed in $\mathcal{I}$\\
%${\bm v}^{\mathcal{B}}$ & Inertial velocity of the quadrotor expressed in $\mathcal{B}$\\
%${\bm v}^{\mathcal{B}}_{{\rm r}}$ & Wind-relative velocity of the quadrotor expressed in $\mathcal{B}$ \\
%${\bm v}^{\mathcal{B}}_{\delta}$ & Wind velocity expressed in $\mathcal{B}$ \\
%${\bm v}^{\mathcal{I}}_{\delta}$ & Wind velocity expressed in $\mathcal{I}$ \\
%$\dot{\bm{x}}$ & inertial velocity vector expressed in $\mathcal{I}$ \\
$\bm{\omega}$ & angular velocity of $\mathcal{B}$ frame with respect to $\mathcal{I}$ frame; $\left[p,q,r\right]^{\rm T}$\\
%$\bm{\dot{\Theta}}$ & Vector containing all Euler angle rates $\left(\dot{\phi},\dot{\theta},\dot{\psi}\right)$\\
$\bm{L}(\bm{\Theta})$ & transformation matrix relating $\bm{\dot{\Theta}}$ and  $\bm{\omega}$ \\ %via a yaw-pitch-roll rotation order\\
%$F_{{\rm D},\left(\cdot\right)}$ & Drag force in the respective $(\cdot)$ body frame direction \\
$C_{{\rm D},\left(\cdot\right)}$ & coefficient of drag in the $(\cdot)$ body frame direction \\
$A_{\left(\cdot\right)}$ & wetted surface area in the $(\cdot)$ body frame direction \\
%$||{\bm v}_{{\rm r}}||$ & Magnitude of the body-frame velocity vector \\
$m$ & quadrotor mass \\
%$g$ & gravity \\
$f_{\left(\cdot\right)}$ & force summation in the respective $(\cdot)$ body frame direction \\
$\tau_{\left(\cdot\right)}$ & moment summation about the respective $(\cdot)$ body frame axis \\
$\bm{u}$ & control input vector; $[T_{\text f}, T_{\text r}]^{\text T}$ \\
%$\dot{\bm{x}}$ & Newton-Euler equation representing inertial velocity \\
%$\dot{\bm{\Theta}}$ & Newton-Euler equation representing inertial rotation rates \\
%$m\dot{\bm{v}}_{\text{r}}$ & Newton-Euler equation representing relative acceleration \\
$\bm{I}$ & vehicle mass moment of inertia matrix\\
$I_{22}$ & pitch-axis mass moment of inertia \\
$\bm{f}_{\text{net}}$ & vector of net body-frame forces; $[f_1,f_2,f_3]^{\rm T}$\\
$\bm{\tau}_{\text{net}}$ & vector of net body-frame moments; $[\tau_1,\tau_2,\tau_3]^{\rm T}$\\
$p_{{\rm N}}, p_{{\rm D}}$ & position north and down, respectively\\
$\delta_{{\rm N}}, \delta_{{\rm D}}$ & wind velocity north and down, respectively \\
$t_{k}$ & discretized time at $k$th time step\\
$u_{{\rm r}}, v_{{\rm r}}, w_{{\rm r}}$ & flow-relative velocity expressed along  $\mathcal{B}$ frame unit vectors ${\bm b}_1$, ${\bm b}_2$, and ${\bm b}_3$, respectively \\
$\bm{x}_{k}$ &  vector of quadrotor states at time $t_{k}$; $\left[p_{{\rm N},{k}},p_{{\rm D},{k}},\theta_{k},u_{{\rm r},{k}},w_{{\rm r},{k}},q_{k}\right]^{\rm{T}}$  \\
$P$ & translating origin of the wind frame \\
$\mathcal{P}$ & wind reference frame; $\{ \mathcal{P}, \bm{p}_{1},{\bm p}_{2},{\bm p}_{3} \} $ \\
%$\bm{p}_{1},{\bm p}_{2},{\bm p}_{3}$ & orthonormal unit vectors of wind frame \\
%in the same orientation as $\bm{i}_{1-3}$, but centered at $P$\\ 
$c$ & speed of the convecting wind-field\\% in the $-\bm{i}_{1}$ direction \\
$\beta$ & horizontal coordinate in $\mathcal{P}$ frame\\
%$\delta_{{\rm N}/D}^{\mathcal{P}}$ & Wind velocity in the north and down directions of the wind frame \\
$\mu(\beta)$ & mean function of the Gaussian process \\
$ \kappa(\beta,\beta')$ & kernel function of the Gaussian process \\
$\mathbb{E}[\cdot]$ & expected value operator \\
$\bm{\theta}$ & vector of Gaussian process hyperparameters; $[L,\sigma]^{\rm T}$  \\
$L$ & length scale of Gaussian process \\
$\sigma^2$ & variance of Gaussian process \\
$h$ & spatial lag term for squared exponential kernel \\ %$\beta - \beta'$ 
%$\text{exp}(\cdot)$ & Exponential operator $e^{(\cdot)}$ \\
$t_m$ & upper bound on the total anticipated takeoff duration \\
$d$ & horizontal length of the operating region  \\
$\lambda$ & horizontal length over which the GP wind-field is estimated \\ % $d + ct_{m}$ 
$A$ & number of anemometers \\
$z_{i}$ & inertial frame horizontal position of the $i$th anemometers \\ %$i \in \{1,\dots,A\}$\\
$\epsilon$ & zero-mean Gaussian measurement noise  with variance $\sigma_{n}^{2}$\\
%$y_{i}(t_k)$ & measurement of $i$th anemometer at time $t_k$ \\
$\bm{y}_k$ & vector of measurements taken by all anemometers at time $t_k$ \\
$\bm{s}_k$ & vector of positions of anemometers for each set of measurements at time $t_k$\\
$F$ & sampling rate of anemometers \\
$t_{N}$ & total sampling time of anemometers prior to takeoff \\
%$J(\cdot)$ & cost functional\\
%$t_{0/f}$ & initial and final times of quadrotor mission \\
%$\left[\cdot\right]_{\text{min/max}}$ & minimum and maximum values of boundary conditions\\
$M$ & total number of measurements used in GP regression across all anemometers and time \\
$\bm{f}$ & vector of wind-field observations \\ %at times spanning $\{t_1,\dots,t_M\}$\\
$\bm{\beta}$ & vector of wind-frame locations corresponding to $\bm{f}$ \\
 %spanning $\{t_1,\dots,t_M \}$\\
$\bm{g}$ & vector of grid points in the wind frame over which the Gaussian process is estimated\\
$G$ & total number of grid points ${\bm g}$   \\
${\bm Y}$ & matrix of wind-sensing time, position, and measurement data collected up to time $t_{N}$\\
$\hat{\delta}_{{\rm N}}(\bm{g};\bm{Y})$ & wind estimate over $\bm{g}$  given data ${\bm Y}$\\
$\bm{P}_{{\delta}_{{\rm N}}}(\bm{g};\bm{Y})$ & covariance matrix of the estimate over $\bm{g}$ given data ${\bm Y}$ \\
$\bm{K}(\cdot,\cdot)$ & %$(M + 1) \times (M + 1)$ 
matrix of covariance values relating observation and/or grid points \\ %using the kernel function\\
$\eta$ & measurement truncation parameter \\
$\phi(t)$ & horizontal interval indicating the zone of measurement acceptance at time $t$\\
%$\hat{(\cdot)}$ & estimate \\
$\mu_w$ & mean of the wind-field
\end{longtable*}}

\section{Introduction}
\lettrine{W}{ind} disturbances can adversely affect small unmanned aerial vehicles (UAVs) by reducing  performance, increasing power consumption, and impacting stability---potentially rendering UAVs unable to perform their tasks or leading to dangerous collisions with people or property \citep{giersch2022atmospheric, mohamed2023gusts}. Feedback control strategies for multirotor-type UAVs to reject disturbance are well developed \cite{alexis2016robust,craig2020geometric,lee2012robust,simon2023flowdrone} and can be aided by measuring the wind (e.g., using hot-wire or sonic anemometers \citep{Hollenbeck.ICUAS.2018, BaileyEtAl.ESSD.2020, BrewerClements.Fire.2020} and multi-hole pressure probes \citep{AlGhussain.AMTD.2020, Yeo.JGCD.2018}) or by inferring the wind using on-board sensors such as a GPS, IMU, and altimeter \citep{McConville.SciTech.2022, PalomakiEtAl.JAOE.2017, Cassano.ESSD.2016, Allison.AST.2020, LangelaanEtAl.GNC.2010}. Model-based wind estimation techniques have also been proposed \citep{GonzalezEtAl.JGCD.2019, Shastry.CSL.2021, Borup.IFAC.2016, Waslander.AIAA.2009}.
While disturbance rejection offers a \emph{reactive} mechanism to mitigate wind already affecting the vehicle,  trajectory planning can avoid or exploit anticipated disturbances at future vehicle positions and thereby provide a \emph{proactive} wind mitigation strategy to complement feedback control. 

Trajectory planning in wind requires a model or estimate of spatiotemporal wind conditions within the operating environment. Spatial wind distributions can be modeled using computational, analytical, and frequency-spectrum models, or by interpolating point measurements provided by wind-sensing infrastructure (e.g., anemometers mounted on structures or other airborne platforms).  Numerical weather and wind prediction models that are routinely used at higher altitudes and around airports have inadequate resolution and accuracy in urban environments to support small UAV flight planning \cite{Glasheen.JAIS.2020, Clark.LincolnLab.2017}.  However, some commercial providers \citep{trueweather_web,gianfelice2022real} offer instrumentation and modeling capabilities for micro-scale wind-field estimation.
%for UAVs in complex urban flows. 
%Sensor placement to optimize such wind-field monitoring has been investigated \citep{PIRHALLA2020117871,GAO2023110803}. %Numerical models of wind flows in urban environments often use 
%Computational methods for wind-field prediction for 
Computational fluid dynamics (CFD) simulations provide numerical models of wind-fields, 
%to predict wind-fields in urban environments
for example, via machine learning from CFD data-sets \cite{vuppala2022wind} or generalizing pre-computed CFD data-sets to different building morphologies for predicting urban wind-fields \cite{galway2012development}. Indeed, prior work has used CFD-based wind models for flight simulation \cite{Davoudi2020, Xue.JAIS.2021, Cybyk.JAIS.2014, Galway.JA.2011}  and path planning \citep{ware2016analysis,patrikar2020real,orr2005framework,raza2017experimental,Baskar.WindDatSet.2021}. While CFD provides high fidelity simulations of complex wind flows, it is computationally intensive. Analytical or statistical wind and gust models are more amenable for use in real-time estimation, control, and trajectory planning.
%Computational studies of urban wind fields have demonstrated that turbulent wind conditions can significantly impact the safety of drone operations \citep{giersch2022atmospheric}, especially near buildings where gusts and wakes dominate \citep{mohamed2023gusts}.  Optimal wind sensor placement in urban environments is a well studied issue, but mainly to identify pollution flow and general urban wind characteristics, and to aid in wind engineering efforts \citep{PIRHALLA2020117871,GAO2023110803}. 
For example,  polynomial \cite{Langelaan.ICRA.2012} or logarithmic  \cite{Rodriguez.IROS.2016} functions of altitude, including those with unsteady components driven by colored-noise \cite{Luders.InfoTech.2013}, have been used to model wind conditions. Other examples of analytical wind and gust model include the one-minus-cosine gust model, the power law used to model wind shear, and various wake vortex \cite{tian2021wind} and parametric models based on potential flow theory \cite{Sydney2013}. Spectral wind turbulence models such as the Dryden and Von Karman models \citep{MIL1797} are widely used for flight dynamics simulations and describe the turbulence characteristics of an aircraft moving at a fixed speed through a spatially ``frozen'' wind-field \cite{beal1993digital,Hess.JGCD.1995}. Other models in this category include the random walk model \cite{tian2021wind} and stochastic Wiener process \cite{Anderson2013}. Interpolation-based methods have been adopted to model wind-fields and ocean currents using B-splines \cite{Dalmau2020}, Kalman filters with spatial basis functions \cite{Peterson.GNC.2011, LanEtAl.TR.2016}, and Gaussian processes \cite{Lee.ICRA.2019, Yang.ACC.2017, Hollinger.JFR.2015, Lawrance.ICRA.2011}.

This work considers a single quadrotor modeled with rigid-body dynamics that operates in the vertical plane in the presence of an uncertain one-dimensional wind-field. The wind-field is vertically uniform, but varies horizontally according to a Gaussian process (GP) model with an unknown mean.  The wind-field propagates through the environment at a constant known speed, and is sampled by several anemometers positioned up-stream from the initial position of the quadrotor.  Local wind  measurements are shared to collaboratively estimate the global wind-field using Gaussian process regression (a form of spatial interpolation). The resulting estimate is treated as a known time and state-dependent disturbance in formulating a deterministic optimal trajectory planning problem for the quadrotor to take off and reach a desired waypoint in the vertical plane. This problem formulation emulates a scenario wherein a quadrotor exploits information provided by nearby wind-sensing infrastructure. 
For example, anemometers might be available to sense wind around an airfield or landing pad, or wind information could be provided by other nearby airborne platforms.
%Wind information might also be available from other airborne platforms operating in the vicinity.   
%The resulting wind estimate is then exploited for nonlinear model-predictive control (NLMPC). The contributions of this paper are: (1) a GP-based estimation framework for multiple quadrotors to collaboratively estimate an unknown and convecting wind-field, and (2) a nonlinear model-predictive control design that incorporates wind predictions to minimize station keeping errors.   The performance of the estimation and control framework is compared through simulations to NLMPC with only local anemometer measurements  and NLMPC with perfect wind knowledge.
%The resulting wind estimate is then provided to a numerical optimal control solver, which calculates the minimum-time trajectory, and generates a set of controls which can be used on a quadrotor in the true wind-field.  In this work, a set of simulations is used to evaluate the proposed control framework.  In these simulations, the numerical optimal control solver, GPOPS-II \citep{gpops} is used to calculate the minimum-time trajectory for a quadrotor to fly from an initial point to a waypoint, using estimated wind-field data.  GPOPS-II provides many optimal control problems as examples, making it easy to implement.  Additionally, it does not parameter tuning to obtain valid results, unlike other control optimization solvers such as linear quadratic regulators (LQR) or the various forms of model predictive controllers (MPC) \citep{Pereida2018,Papachristos2015}.

The contributions of this paper are: (1) a  GP-based estimation framework for using noisy wind-sensing infrastructure in an environment to collaboratively estimate a one-dimensional wind-field that is convecting at a known speed, and (2) formulating a minimum-time takeoff trajectory planning problem for a quadrotor with vertical plane rigid-body dynamics that  incorporates estimated wind knowledge. The trajectory planning problem is solved using the pesudeospectral optimal control solver GPOPS-II \citep{gpops}. The performance of the estimation and trajectory planning framework is compared through simulations that vary the wind strength and variance, and  the measurement frequency and noise.  
%Numerical optimal control solutions generated with estimated wind-field information are compared against solutions calculated using the generated optimal control but in the true wind-field. 

The remainder of the paper is organized as follows. Section~\ref{sec:problem_formulation} describes the quadrotor motion model, the wind and wind measurement models, and the optimization problem. Section~\ref{sec:estimation_control} describes a Gaussian process approach for estimating the wind-field and a GPOPS-II based trajectory planning. Section~\ref{sec:results} describes the results of numerical simulations that quantify control performance under varying wind conditions and sensor quality. The paper is concluded, and future work is suggested in Section~\ref{sec:conclusion}.

\section{Problem Formulation}\label{sec:problem_formulation} 
This section introduces the vertical plane dynamics of a quadrotor, presents a model for wind-sensing infrastructure measurement, and formulates the optimization problem.

\subsection{Quadrotor Dynamics}\label{sec:quad_dynamics}
\begin{figure}[H]
    \centering
    \includegraphics[width=0.4\textwidth]{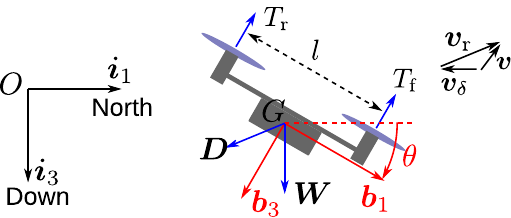}
    \caption{References frames and quantities used to define the three degree-of-freedom quadrotor model. The wind triangle (upper right) shows the inertial velocity ${\bm v}$ as the sum of the flow-relative velocity ${\bm v}_{{\rm r}}$ and the wind velocity ${\bm v}_\delta$. The quadrotor is drawn with pitch angle $\theta < 0$.}
    %The drag force ${\bm D}$ opposes the air-relative velocity ${\bm v}_{{\rm r}}$.}
    \label{fig:quad}
\end{figure}

Let $\mathcal{I} = \{O, \bm{i}_{1}, \bm{i}_{2}, \bm{i}_{3} \}$ be an inertial reference frame with its origin at point $O$ and with orthonormal unit vectors oriented along the north-east-down directions, respectively, as shown in Fig.~\ref{fig:quad}.  Let $\mathcal{B} = \{G, {\bm{b}}_{1}, {\bm{b}}_{2}, {\bm{b}}_{3} \}$ be a body reference frame, with its origin centered at the center of mass ($G$) of the quadrotor, and with orthonormal unit vectors oriented in the forward-right-down body directions, respectively.  

\noindent Let $\bm{\Theta} = \left[\psi, \theta, \phi\right]^{\text T}$ be a vector of yaw, pitch, and roll angles respectively. The rotation matrix
\begin{equation}\label{eqn:rot}
    {\bm R}({\bm \Theta})= \begin{bmatrix}
c_\theta c_\psi & s_\phi s_\theta c_\psi - c_\phi s_\psi & c_\phi s_\theta c_\psi + s_\phi s_\psi  \\
c_\theta s_\psi & s_\phi s_\theta s_\psi + c_\phi c_\psi & c_\phi s_\theta s_\psi - s_\phi c_\psi \\
- s_\theta & s_\phi c_\theta & c_\phi c_\theta
    \end{bmatrix}
\end{equation}
is an element of special orthogonal group, ${\rm SO}(3)$, and relates the orientation of reference frame $\mathcal{B}$ to $\mathcal{I}$ using a 3-2-1 Euler angle sequence, where $c(\cdot) = \cos{}$, $s(\cdot) = \sin{}$, are used as shorthand notation.  The inertial position of the quadrotor is $\bm{x} = \left[ p_{\text{N}}, p_{\text{E}}, p_{\text{D}}\right]^{\text T}$ expressed using coordinates in $\mathcal{I}$ and the inertial velocity of the quadrotor is ${\bm v}^{\mathcal{B}} = \left[u,v,w\right]^{\text T}$ expressed  in $\mathcal{B}$. The inertial velocity is given by the sum 
\begin{equation}
{\bm v}^{\mathcal{B}} = {\bm v}_{{\rm r}}^{\mathcal{B}} + {\bm v}_\delta^{\mathcal{B}} \label{eq:vel_triangle}\;,
\end{equation}
where ${\bm v}_{{\rm r}}^{\mathcal{B}} = \left[u_{{\rm r}},v_{{\rm r}},w_{{\rm r}}\right]^{\text T}$ is the wind-relative velocity of the quadrotor and ${\bm v}_{\delta}^{\mathcal{B}} = [\delta_1, \delta_2, \delta_3]^{\text T}$ is the wind velocity, both expressed in $\mathcal{B}$.  The inertial velocity in frame $\mathcal{I}$ is then $\dot {\bm x} = {\bm v}^\mathcal{I}$ = ${\bm R}({\bm \Theta}){\bm v}^{\mathcal{B}}$.
Similarly, the wind-velocity in the inertial frame is ${\bm v}^\mathcal{I}_\delta = [\delta_{{\rm N}},\delta_{\rm{E}},~\delta_{{\rm D}} ]^{\text T}$ and is related to the body-frame wind-velocity by
\begin{align}
{\bm v}^\mathcal{I}_\delta &= {\bm R}({\bm \Theta}){\bm v}^\mathcal{B}_\delta \;.
\label{eq:wind_inertial_body_relationship}
\end{align}
%By using the rotation defined by (\ref{eqn:rot}), the translational dynamics can readily be defined in $\mathcal{I}$ using the components of velocity in $\mathcal{B}$ (i.e. $\bm{\dot{x}}_{\mathcal{B}} = \left[u,v,w\right]^{\text T}$).  
The vector $\bm{\omega} = \left[p,q,r\right]^{\text T}$ is the angular velocity of the quadrotor body-frame with respect to the inertial frame, where $p$, $q$, and $r$ are the roll, pitch, and yaw rates,  respectively.  
% Using the aforementioned 3-2-1 Euler angle sequence, the Euler rates $\bm{\dot{\Theta}} = \left[\dot{\phi},\dot{\theta},\dot{\psi}\right]^{\text T}$ can be converted to the body frame angular velocity $\bm{\omega} = \begin{bmatrix} p & q & r \end{bmatrix}^T$ via the transformation \citep{BeardMcLain.UAVbook} 
% \begin{align}\label{eq:angVelRot}
%     \underbrace{
%     \begin{bmatrix}
%     p \\ q \\ r
%     \end{bmatrix}}_{\bm{\omega}} &= \begin{bmatrix}\dot{\phi} \\ 0 \\ 0\end{bmatrix} + \bm{R}(\left[\phi, 0, 0\right]^T) \begin{bmatrix}0 \\ \dot{\theta}\\ 0\end{bmatrix} + \bm{R}(\left[\phi, 0, 0\right]^T) \bm{R}(\left[0, \theta, 0\right]^T)\begin{bmatrix}0 \\ 0\\ \dot{\psi}\end{bmatrix}\\
%     &= \underbrace{\begin{bmatrix}
%     1 & 0 & -s_{\theta} \\
%     0 & c_{\phi} & s_{\phi}c_{\theta}\\
%     0 & -s_{\phi} & c_{\phi}c_{\theta}
%     \end{bmatrix}}_{\bm{L}(\bm{\Theta})} \underbrace{\begin{bmatrix}\dot{\phi} \\ \dot{\theta}\\ \dot{\psi}\end{bmatrix}}_{\dot{\bm{\Theta}}} \;.
% \end{align}
The Euler rates $\dot{\bm{\Theta}} = \left[\dot{\phi},\dot{\theta},\dot{\psi}\right]^{\rm{T}}$ are related to $\bm{\omega}$ by $\dot{\bm{\Theta}} = \bm{L}(\bm{\Theta}) \bm{\omega}$ where 
\begin{equation}\label{eqn:intermediaterotation}
    {\bm L}({\bm \Theta})  = \begin{bmatrix}
        1 & s_{\phi}t_{\theta} & c_{\phi}t_{\theta} \\
        0 & c_{\phi} & -s_{\phi} \\
        0 & {s_{\phi}}/{c_{\theta}} & {c_{\phi}}/{c_{\theta}}
    \end{bmatrix} \;.
\end{equation}
The Newton-Euler equations of motion for the system are given by \citep{GonzalezEtAl.JGCD.2019,BeardMcLain.UAVbook}
\begin{align}
    \dot{\bm{x}} &= \bm{R}(\bm{\Theta}) (\bm{v}_{\text{r}} + \bm{v}_{\delta}) \label{eqn:p_dot}\\
    \dot{\bm{\Theta}} &= \bm{L}(\bm{\Theta}) \bm{\omega}  \label{eqn:theta_dot} \\
    m\dot{\bm{v}}_{\text{r}} &=m\bm{v}_{\text{r}} \times \bm{\omega} + \bm{f}_{\text{net}}  \label{eqn:v_dot} \\
    \bm{I} \dot{\bm{\omega}} &=\bm{I} \bm{\omega} \times  \bm{\omega} +\bm{\tau}_{\text{net}} \;, \label{eqn:omega_dot} 
\end{align}
where ${\bm I} \in \mathbb{R}^{3 \times 3}$ is the inertia matrix, ${\bm f}_{\rm net} = [f_1, f_2, f_3]^{\rm T}$, and ${\bm \tau}_{\rm net} = [\tau_1, \tau_2, \tau_3]^{\text T}$ are the net body-frame forces and moments acting on the vehicle, respectively.

In this work, the six degree of freedom system \eqref{eqn:p_dot}--\eqref{eqn:omega_dot} is simplified to a three degree of freedom model of the quadrotor's longitudinal dynamics, consisting of the north-down positions ($p_{\text{N}}$, $p_{\text{D}}$) and the pitch angle $\theta$ (i.e., ignoring $p_{\rm{E}}$ and assuming $\psi = \phi = v_{{\rm r}} = p = r =0$).  The control forces acting on the quadrotor are the front thrust $0 \leq T_{\rm f}  \leq T_{\rm max}$ and rear thrust $0 \leq T_{\text r} \leq T_{\rm max}$ that are aligned with the $-{\bm b}_3$ direction where $T_{\rm max}$ is the maximum thrust. Gravity $g$ is aligned with the ${\bm i}_3$ direction, and a low-speed quadratic drag acts in the direction opposite to ${\bm v}_{{\rm r}}$. Drag components are modeled as 
\begin{align}
    % F_{{\rm D},1} &= -\frac{1}{2} \rho C_{{\rm D},1} A_1 ||{\bm v}_{{\rm r}}|| u_{{\rm r}} \label{eq:fd1} \\
    % F_{{\rm D},3} &= -\frac{1}{2} \rho C_{{\rm D},3} A_3 ||{\bm v}_{{\rm r}}|| w_{{\rm r}} \label{eq:fd2} \;,
         F_{{\rm D},1} &= -\frac{1}{2} \rho C_{{\rm D},1} A_1  u_{{\rm r}}^2~{\rm sign}(u_{\rm r})  \label{eq:fd1} \\
    F_{{\rm D},3} &= -\frac{1}{2} \rho C_{{\rm D},3} A_3  w_{{\rm r}}^2~{\rm sign}(w_{\rm r}) \label{eq:fd2} \;,
\end{align}
where 
%$||\bm{v}_{{\rm r}}||$ is the magnitude of the body-frame flow-relative velocity vector, and 
the constants $(C_{{\rm D},1}, C_{{\rm D},3})$ and $(A_1, A_3)$ are the coefficients of drag and the projected surface areas in the ${\bm b}_1$ and ${\bm b}_3$ directions, respectively.  
%Note that the vector $||\bm{v}_{{\rm r}}||\cdot [u_{{\rm r}}, w_{{\rm r}}]^{\text T}$ in \eqref{eq:fd1}--\eqref{eq:fd2} has  magnitude $||\bm{v}_{{\rm r}}||^2$ and is oriented in the direction of the relative velocity. 
%The magnitude of the vehicle velocity is multiplied by the vehicle velocity in each body-frame direction to obtain the squared velocity term of the general drag equation. 
The sum of forces and moments along the body-frame axes are
\begin{align}
f_1 &=  - mg\sin \theta  + F_{{\rm D},1} \\
f_3 &=  mg\cos \theta  + F_{{\rm D},3} - T_{\text f} - T_{\text r}\\
\tau_2 &= (T_{\text f} - T_{\text r})l \;.
\end{align}
The control input is $\bm{u} = [T_{\text f}, T_{\text r}]^{\text T}$.  
%The dynamics \ref{eqn:p_dot}-\ref{eqn:omega_dot} include the wind velocity gradient term ${\bm \Phi}={\bm R}^T(\partial {\bm v}_\delta/\partial {x})^{\text T}{\bm R}$ and  the body-frame wind velocity ${\bm v}^B_{\text w}  = {\bm R}({\bm \Theta}){\bm v}_\delta({x})$ which enters as a translational disturbance in \eqref{eqn:p_dot}. 
%The wind-induced angular velocity ${\bm \omega}_{\text w}$ will be evaluated using the velocity point method \citep{JAIS} (i.e., comparing velocity at select  points along the body). Note that ${\bm \omega} = {\bm \omega}_{\text r} + {\bm \omega}_{\text w}$ and thus the computed ${\bm \omega}_{\text w}$ appears in \eqref{eqn:theta_dot}-\eqref{eqn:omega_dot}.
% \begin{align}
%  {\bm f}_{\text net} &= {\bm f}_{\text aero}({\bm v}_{{\rm r}}) + {\bm f}_{\text gravity}({\bm \Theta}) + {\bm f}_{\text control}  \\
%   {\bm \tau}_{\text net} &=  {\bm \tau}_{\text aero}({\bm v}_{{\rm r}}, {\bm \omega}_{{\rm r}}) +{\bm \tau}_{\text control} 
% \end{align}
Under the simplified dynamics  \eqref{eq:wind_inertial_body_relationship} becomes
\begin{align}
\delta_{{\rm N}} &= \cos \theta \delta_1 + \sin \theta \delta_3 \\
\delta_{{\rm D}} &= -\sin \theta \delta_1 + \cos \theta \delta_3
\end{align}
% \begin{equation}
%     %\dot{x} = 
%     \begin{bmatrix}
%         \dot{p}_{{\rm N}} \\ \dot{p}_{{\rm D}} \\ \dot{\theta} \\ \dot{u}_{{\rm r}} \\ \dot{w}_{{\rm r}}  \\ \dot{q}
%     \end{bmatrix} = \begin{bmatrix}
%         (u_{{\rm r}}+\delta_1)\cos{\theta} + (w_{{\rm r}}+\delta_3)\sin{\theta} \\ -(u_{{\rm r}}+\delta_1)\sin{\theta}+(w_{{\rm r}}+\delta_3)\cos{\theta} \\ q  \\ -qw_{{\rm r}} \\ qu_{{\rm r}} \\ 0
%     \end{bmatrix} + \begin{bmatrix}
%         0 \\ 0  \\ 0 \\ f_1/m \\ {f_3}/m \\ \tau_2/I_{22}
%     \end{bmatrix}
%     \label{eq:simpEOM}
% \end{equation}
% where $I_{22}$ is the mass moment of inertia of the quadrotor around its pitch axis. Under the simplified dynamics, $\delta_2 = \delta_{\rm{E}} = 0$  and \eqref{eq:wind_inertial_body_relationship} becomes:
% \begin{align}
% \delta_{{\rm N}} &= \cos \theta \delta_1 + \sin \theta \delta_3 \\
% \delta_{{\rm D}} &= -\sin \theta \delta_1 + \cos \theta \delta_3
% \end{align}
% so that \eqref{eq:simpEOM} can be rewritten as
and equations \eqref{eqn:p_dot}--\eqref{eqn:omega_dot} simplify to:
\begin{equation}
    %\dot{x} = 
    \begin{bmatrix}
        \dot{p}_{{\rm N}} \\ \dot{p}_{{\rm D}} \\ \dot{\theta} \\ \dot{u}_{{\rm r}} \\ \dot{w}_{{\rm r}}  \\ \dot{q}
    \end{bmatrix} = \begin{bmatrix}
        u_{{\rm r}} \cos{\theta} + w_{{\rm r}} \sin{\theta} + \delta_{{\rm N}} \\ -u_{{\rm r}}\sin{\theta}+w_{{\rm r}}\cos{\theta}  \\ q \\ -qw_{{\rm r}} \\ qu_{{\rm r}} \\ 0
    \end{bmatrix} + \begin{bmatrix}
        0 \\ 0  \\ 0 \\ f_1/m \\ {f_3}/m \\ \tau_2/I_{22}
    \end{bmatrix}= 
    \begin{bmatrix}
        \text{Position north} \\
        \text{Position down} \\
        \text{Pitch angle} \\
        \text{Flow-relative velocity } \bm{b}_1 \\
        \text{Flow-relative velocity } \bm{b}_3\\
        \text{Pitch rate}
    \end{bmatrix}
    \label{eq:simpEOM}
\end{equation}
where $I_{22}$ is the pitch-axis mass moment of inertia of the quadrotor and the wind vertical component has been set to zero ($\delta_{{\rm D}} = 0)$.  The state vector denoting the location, pitch, and velocities of the quadrotor at a particular time instant $t_{k}$ is $\bm{x}_{k} = \left[p_{{\rm N},{k}},p_{{\rm D},{k}},\theta_{k},u_{{\rm r},{k}},w_{{\rm r},{k}},q_{k}\right]^{\rm{T}}$.

% \subsection{Anemometer placement}
% An environment is presented in which there are several single-axis anemometers mounted up-stream of the wind-field from the initial position of the quadrotor.  The anemometers are assumed to be mounted perpendicular to the ground plane, with their primary measurement axes aligned in the direction the wind-field is travelling.  Because the wind-field only varies horizontally, only the horizontal position of the anemometers must be defined.  Let $p_{{\rm N}_{0}}$ be the initial position of the quadrotor in the North direction of the inertial frame.  Further, let $\delta$ be the distance of the first anemometer from this initial position, and let $\gamma$ be the distance from the first anemometer to the last in the simulation environment.  Define $M$ to be the number of anemometers, and define $s_i$ as the position in the inertial frame of anemometer $i$ as calculated by

% \begin{equation}\label{eqn:anemPlacement}
%     s_i = p_{{\rm N}_{0}} + \frac{\gamma(i-1)}{M} \forall i\in\{1,\cdots,M\}
% \end{equation}.

\subsection{Convected Wind Model}\label{sec:wind_model}
\begin{figure}[ht!]
    \centering
    \includegraphics[width=0.5\columnwidth]{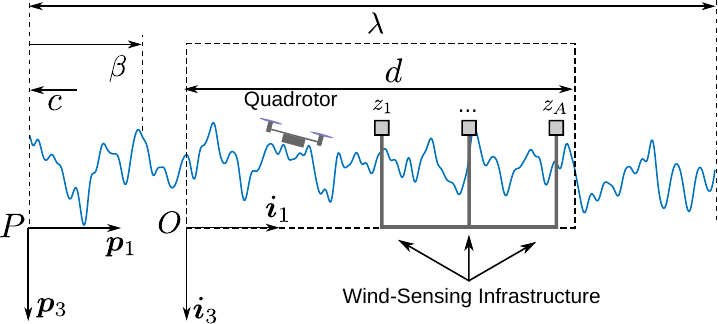}
    \caption{Wind-field model where coordinate $\beta$ represents the distance from the origin $P$ in the wind frame, $c$ is the wind propagation speed at which the origin $P$ moves to the left, and $d$ is the operating region of the quadrotor.  The wind-profile used for simulation (and estimation) has length $\lambda = d +ct_m$, and the horizontal coordinate $\beta \in [0, \lambda]$ is the distance from $P$ in the ${\bm p}_1$ direction.  
    %Based on the selected number of anemometers, they are spaced evenly within these bounds.
    }
    \label{fig:multi_quad}
\end{figure}

The wind is modeled as uniform with altitude and propagating horizontally, as shown in Fig.~\ref{fig:multi_quad}. Let the wind-field in the inertial frame be denoted ${\bm v}^\mathcal{I}_\delta \left(p_N,t\right)$  
where
%where $x$ is a coordinate along the $ {\bm i}_1$ axis and 
$t \geq 0$ is time. 
The wind-profile appears``frozen'' when viewed in a translating wind frame $\mathcal{P} = \{P, {\bm p}_1, {\bm p}_2, {\bm p}_3\}$ that initially has its origin $P$ coinciding with the inertial frame origin $O$ at time $t = 0$ and moves along $-{\bm i}_1$ with a known velocity $c$.  Let $\beta$ denote the horizontal coordinate in $\mathcal{P}$. The wind-field is fully specified by 
$\delta_{{\rm N}}^\mathcal{P}(\beta)$
%${\bm v}^\mathcal{P}_\delta \left(\beta \right)$ 
and relates to the inertial wind-field by 
\begin{equation}\label{eqn:wind}
    \delta_{{\rm N}}^\mathcal{I} (p_{{\rm N}},t) = \delta_{{\rm N}}^\mathcal{P} \left(p_{{\rm N}} + ct \right)\;,
\end{equation}
where $ c < 0$ and $\beta = p_{{\rm N}} + ct$. 
%the $\mathcal{P}$ frame and $\mathcal{I}$ horizontal coordinates . 
This paper adopts a Gaussian process (GP) model for $\delta_{{\rm N}}^\mathcal{P} \left(\beta \right)$; however, in principle, other  wind profiles $\delta_{{\rm N}}^\mathcal{P} \left(\beta \right)$ can be estimated using GP regression.

The wind profile is a scalar, one-dimensional GP, $\delta_{{\rm N}}^\mathcal{P} \left(\beta \right)$, which is a {random function} of an argument $\beta \in \mathbb{R}$ representing the horizontal position relative to the origin of the wind frame $\mathcal{P}$. The GP is completely specified by its mean function $\mu(\beta)$ over the input space and covariance function $\kappa({\beta},{\beta}')$ (also called the \emph{kernel}). 
% Let $\mu(\beta)$ denote the mean over the input space and let $\kappa({\beta},{\beta}')$ denote the covariance function (also called the \emph{kernel}). Formally, 
\begin{align}
\mu(\beta) &= \mathbb{E}[\delta_{{\rm N}}^\mathcal{P} \left(\beta \right)] \label{eq:gp_mean} \\
\kappa({\beta},{\beta}') &= \mathbb{E}[\{\delta_{{\rm N}}^\mathcal{P} \left(\beta \right)-\mu(\beta)\}\{\delta_{{\rm N}}^\mathcal{P}\left(\beta' \right)-\mu(\beta)') \}] \label{eq:gp_cov} \;,
%= {\text Cov}({x}, {x}')   
\end{align}
\noindent where $\mathbb{E}[\cdot]$ is the expected value operator.
%The shape of the covariance function $\kappa({x},{x}')$ is typically parameterized by a scale and other constants called 
%\emph{hyperparameters}. 
This work considers the squared exponential kernel
\begin{equation}
\kappa({h}; \bm \theta) = \sigma^2 \text{exp}\left( -\frac{{h}^2}{2L^2}\right) ,
\label{eq:kernel}
\end{equation}
where ${h} = \beta - \beta'$ is the spatial lag term, and ${\bm \theta} = [L, \sigma]^{\text T}$ is a vector of hyperparameters where $L$ is the length scale of the GP and $\sigma^2$ is the variance. The mean $\mu(\beta)$ is assumed to be constant. The length scale relates to the smoothness of the wind-field over the spatial coordinate $\beta$ (i.e., gust duration), whereas the variance relates to the amplitude of peaks and troughs (i.e., gust magnitude). A realization of a GP can be generated, for example, by drawing a Gaussian random vector (representing the value of the realization over a grid of points) from a mean and covariance matrix defined according to \eqref{eq:gp_mean}--\eqref{eq:gp_cov}. In this work, realizations of the GP wind-profile are generated for uniformly spaced grid points over a distance $\lambda = d+ct_m $, where $d$ is the horizontal region within which the quadrotor and anemometers operate, and $t_m$ is an upper bound on the total anticipated sensing and  takeoff duration. The span $\lambda$ allows for the profile to be well-defined within the operating region for all $ t \in [0, t_m]$. 
%and is also the horizontal length over which the wind-field is estimated prior to takeoff.
%The wind-field is assumed to be a Gaussian process that is parameterized by the hyperparameters $L$ and $\sigma$ for the spatial length scale and spatial variance, respectively.  The area within which the quadrotor and anemometers operate is defined as the horizontal region $d$ with a total span of $h = d+ct_m$, where $t_m$ is the total anticipated takeoff duration.  The correlation of any two points defined over $h$ is defined by the aforementioned hyperparameters via a covariance function, as further detailed in Sec. \ref{sec:gp}.  

% We define $B$ as the grid of coordinates along ${\bm p}_1$ on which the wind speed is specified by ${\bm v}^\mathcal{P}_\delta$.  Given the known translational velocity of the wind-field $c$, the Matlab linear interpolation function, \emph{interp1}, was used to obtain a local measurement of the wind speed, at the position of each anemometer \cite{interp1}.  A zero-mean, Gaussian measurement noise $\epsilon \sim \mathcal{N}(0,\sigma_{n}^{2})$, with variance $\sigma_n^2$ was included to simulate the noise associated with taking wind measurements using an anemometer.  Using the Matlab linear interpolation function to linearly interpolate the wind-field from the position of anemometer $i$, the wind measurements provided by each anemometer are defined as

% \begin{equation}\label{eq:anemMsmt}
%     y_i = \text{\emph{interp1}}(B, {\bm v}^\mathcal{P}_\delta, s_i + ct_k) + \epsilon
% \end{equation}

% \noindent where $t_k$ is the discrete simulation time during which the measurement is taken, and $s_i + ct_k$ is the time shifted position of anemometer $i$ in the wind frame.

\subsection{Wind Sensing Infrastructure}\label{sec:commModel}
The operating environment of the quadrotors is instrumented with a total of  $A$ networked wind-sensing instruments (anemometers) that are positioned upstream of the quadrotor takeoff position at fixed locations $z_i$ for $i = 1, \ldots, A$, measured as the horizontal distance from $O$. 
%(Since the wind-field only varies horizontally, only the horizontal position of the anemometers is required.) 
An anemometer located at position $z_i$ measures
\begin{equation}
    y_i(t_{k}) =  \delta_{{\rm N}}^\mathcal{I}(z_i,t_{k}) + \epsilon
    \label{eq:msmt_noisy}
\end{equation}
at time $t_{k}$, where $\epsilon \sim \mathcal{N}(0,\sigma_{n}^{2})$ is zero-mean, Gaussian measurement noise with variance $\sigma_{n}^{2}$ and $\delta_{{\rm N}}^\mathcal{I}(z_i,t_{k})$ is obtained from \eqref{eqn:wind} assuming a wind profile $\delta_{{\rm N}}^\mathcal{P}$.
%. The anemometers are assumed to be mounted perpendicular to the ground plane, with their primary measurement axes aligned in the direction the wind-field is traveling. Because the wind-field only varies horizontally, only the horizontal position of the anemometers must be defined. The quadrotors are located at horizontal positions $s_i$ for $i = 1, \ldots, M$ within the operating region. For the purposes of this study the wind direction is known and the quadrotors are assumed to be located downstream of the initial position of the quadrotor. 
Define the vector of measurements taken by all anemometers at a discrete-time instant $t_{k}$ as $\bm{y}_{k} = [ {y}_1,  \ldots, {y}_{A}]^T$.  Similarly, define the vector of positions of the anemometers for each set of measurements as $\bm{s}_{k} = [ z_{1}, \ldots, z_{A}]^T$.  Let $F$ be the sampling rate (Hz) of the sensors so that for consecutive sample times $t_k - t_{k-1} = (1/F)$. Prior to takeoff, the anemometers collect data for a total of $t_{N}$ seconds as represented by the matrix
\begin{equation}\label{eq:msmtMatrix}
    \bm{Y} = \begin{bmatrix}
        t_1 & t_2 & \cdots & t_{N} \\ \bm{s}_1 & \bm{s}_2 & \cdots & \bm{s}_{N} \\\bm{y}_1 & \bm{y}_2 & \cdots & \bm{y}_{N}
    \end{bmatrix}^T \in \mathbb{R}^{(2A+1) \times N} \;,
\end{equation}
where $t_1, \ldots, t_{N}$ are the times at which measurements are taken.
%\bm{s}_1, \ldots, \bm{s}_N \in \mathbb{R}^{A}$ are the locations vectors of the anemometers, and $\bm{y}_1, \ldots, \bm{y}_N \in \mathbb{R}^A$ are measurement vectors defining the respective measurements taken by each anemometer. 
The information gathered by the anemometers preceding takeoff is made available to a centralized planner for wind estimation and trajectory planning.

\subsection{Problem Statement}\label{sec:probStatement}
The objective is to plan a minimum-time trajectory to a desired position $(p_{{\rm N},\rm{final}}, p_{{\rm D},\rm{final}})$ for a quadrotor that has access to nearby wind sensing infrastructure in an uncertain wind-field.  Let $t_{\rm init} \geq t_{N}$ be the takeoff time of the quadrotor and $t_{\rm final}$ be the final time that the quadrotor reaches the  desired position. The optimal control problem is to
\begin{equation}
    ~{\rm minimize}\qquad J(\bm{u}(\cdot)) = \int_{t_{\rm{init}}}^{t_{\rm{final}}}dt = t_{\rm{final}} - t_{\rm{init}}
\end{equation}
subject to the boundary conditions
\begin{equation}
\begin{aligned}
p_{{\rm N}}(t_{\rm init}) &= p_{{\rm N},\rm{init}}  \\
p_{{\rm D}}(t_{\rm init}) &= p_{{\rm D},\rm{init}} \\
\theta(t_{\rm init}) &= 0\\
u_{{\rm r}}(t_{\rm init}) &= 0 \\
w_{{\rm r}}(t_{\rm init}) &= 0\\
q(t_{\rm init}) &= 0
\end{aligned}
\qquad  \qquad \qquad 
\begin{aligned}
p_{{\rm N}}(t_{\rm final}) &= p_{{\rm N},\rm{final}}  \\
p_{{\rm D}}(t_{\rm final}) &= p_{{\rm D},\rm{final}} \\
\theta_{\rm{min}} \leq \theta(t_{\rm final}) &\leq \theta_{\rm{max}} \\
u_{{\rm r},\rm{min}} \leq u_{{\rm r}}(t_{\rm final}) &\leq u_{{\rm r},\rm{max}}\\
w_{{\rm r},\rm{min}} \leq w_{{\rm r}}(t_{\rm final}) &\leq w_{{\rm r},\rm{max}} \\
q_{\rm{min}} \leq q(t_{\rm final}) &\leq q_{\rm{max}} 
\end{aligned} \;,
% \begin{aligned}
% p_{{\rm N}}(t_f) &= p_{{\rm N},f}  \\
% p_{{\rm D}}(t_f) &= p_{{\rm D},f} \\
% \theta(t_f) &= \theta_{\rm{min}} \leq \theta(t_f) \leq \theta_{\rm{max}} = -30^{\circ} \leq \theta(t_f) \leq 30^{\circ}\\
% u_{{\rm r}}(t_f) &= u_{{\rm r},\rm{min}} \leq u_{{\rm r}}(t_f) \leq u_{{\rm r},\rm{max}} = -5 \rm{ m/s} \leq u_{{\rm r}}(t_f) \leq 5 \rm{ m/s}\\
% w_{{\rm r}}(t_f) &= w_{{\rm r},\rm{min}} \leq w_{{\rm r}}(t_f) \leq w_{{\rm r},\rm{max}} = -5 \rm{ m/s} \leq w_{{\rm r}}(t_f) \leq 5 \rm{ m/s}\\
% q(t_f) &= q_{\rm{min}} \leq q(t_f) \leq q_{\rm{max}} = -100^{\circ} \rm{/s} \leq q(t_f) \leq 100^{\circ} \rm{/s}
% \end{aligned} \;,
\end{equation}
the dynamics \eqref{eq:simpEOM} and the control constraints $0 \leq T_{\text{f}}, T_{\rm r} \leq T_{\text{max}}$. The term $\delta_{{\rm N}}$ appearing in the dynamics \eqref{eq:simpEOM} is not known and is  estimated using  available data \eqref{eq:msmtMatrix}. The terminal boundary conditions include a specific waypoint in the vertical plane $(p_{{\rm N},\rm{final}}, p_{{\rm D},\rm{final}})$ and inequality constraints for the minimum and maximum pitch angle, wind-relative velocities, and pitch rate.
%from $t_0 = 0 \text{ sec}$ to $t_f$, subject to the boundary conditions $\bm{x}(t_0) = \left[p_{{\rm N},0}, p_{{\rm D},0}, 0, 0, 0, 0\right]$ and \\$\bm{x}(t_f) = \left[p_{{\rm N},f}, p_{{\rm D},f}, \pm \theta_{f}, \pm u_{{\rm r},f}, \pm w_{{\rm r},f}, \pm q_{f}\right]$.  The cost functional is also subject to the thrust limits for the front and rear rotors $0 \leq T_{\text{f}} \leq T_{\text{max}}$ and $0 \leq T_{\text{r}} \leq T_{\text{max}}$ respectively.

\section{Wind Estimation and Trajectory Planning}\label{sec:estimation_control}
% To address the time-optimal control problem of Sec. \ref{sec:probStatement}, and to evaluate the effects of using a wind-field estimation to calculate the control effort of a quadrotor in an uncertain wind-field, the following strategies are proposed.  
This section proposes a GP-based wind estimation procedure for the convected wind-field model and describes how the wind-estimate is used to formulate the minimum-time trajectory planning problem to be solved numerically.
%is performed and subsequently used to formulate the dynamics for use in a numerical optimal control solver.  First, the wind-field estimation is described through an brief explanation on Gaussian process based regression.  Methods are presented which were used to propagate the wind-field over time, and to truncate the data according to the defined operating area.  An illustrative example is shown to display how the estimate evolves through the operating area over time.  Finally, the proposed control framework is presented, which uses the results calculated via a numerical optimal control solver to control the quadrotor.

\subsection{Gaussian Process Wind-field Regression}\label{sec:gp}
The raw data collected by the wind-sensing infrastructure is encapsulated in the matrix ${\bm Y}$ \eqref{eq:msmtMatrix}. The measurements are collected at sampling locations $z_i$ at times $t_i$ for $i = 1,\ldots, N$. However, since the wind-field is convected at speed $c$ the measurement locations in the wind frame are 
\begin{equation}
\beta_i = z_i - ct_i \;. % (t - t_i) \;.
\label{eq:msmt_correction}
\end{equation}
% and in the inertial frame
% \begin{equation}
% \overline{s}_i(t) = s_i - c(t - t_i) \;.
% \label{eq:msmt_correction}
% \end{equation}
%for each anemometer but at different times and hence represent spatially distinct samples
%, they were collected at different times. In a convected wind-field this is equivalent to sampling different points along the wind profile. 
%For each set of measurements prior to the end of the sampling time of the anemometers $(T_{n})$, under the assumption of a propagated wind-field model (Sec.~\ref{sec:wind_model}) with a known, fixed, speed $c$, the locations $s_i$ of the measurements at taken at corresponding times $t_i < t_k$ are shifted to the current time $t_k$ by translating them according to $\overline{s_{i}^{k}} = s_i - c (t_k - t_i)$.  The shifted measurement location $s_{k}$ then represents the updated location of the measurement $y_i$, in the inertial frame.  At each time instant that a new set of measurements $\bm{y}_k$ is taken, the new measurement and the set of shifted prior measurements $\{s_{ik},y_i\}$ are used to construct ${\bm{s}}$ and ${\bm{f}}$ in the GP regression, as shown in \eqref{eq:posterior1}--\eqref{eq:posterior2}.  
%If the convection speed $c$ is unknown it may be replaced with an estimate $\hat c$ or may be ignored entirely with $c = 0$.
%\subsection{Truncation Strategy}
The wind estimation approach is based on ordinary Kriging, which is a form of GP regression that handles GP process models with an unknown constant mean. Given a set of $M$ observations of the wind-field,  ${\bm f} = [{y}_1^{\rm T}, \ldots, {y}_M^{\rm T}]^{\text T} $ at wind-frame locations ${\bm \beta} = [\beta_1, \ldots, \beta_M]^{\text T}$ the ordinary Kriging  estimator \citep[Ch. 4]{Olea1999} predicts the estimate of the wind $\hat{\delta}_{{\rm N}}^\mathcal{P}({\bm g}; {\bm Y})$
%process ${\bm f}^* = [f_1^*, \ldots, f_G^*]^{\text T} $ 
at a vector of  grid points ${\bm g} = [g_1, \ldots, g_G]^{\text T}$ along with the corresponding covariance matrix $ {\bm P}_{{\delta}_{{\rm N}}}({\bm g}; {\bm Y}) \in \mathbb{R}^{G \times G}$. The prediction is computed according to 
\begin{align}
\hat{\delta}_{{\rm N}}^\mathcal{P}({\bm g}; {\bm Y})  %{{\bm f}_*|{\bm f}} 
&= 
\begin{bmatrix}
{\bm f}^{\text T} & 
0
\end{bmatrix}
{\bm K}({\bm \beta},{\bm \beta})^{-1}  {\bm K}({\bm g},{\bm \beta}) 
 \label{eq:estimate} \\
 {\bm P}_{{\delta}_{{\rm N}}}({\bm g}; {\bm Y})&= {\bm K}({\bm g},{\bm g}) - {\bm K}({\bm g},{\bm \beta}) {\bm K}({\bm \beta},{\bm \beta})^{-1}{\bm K}({\bm g},{\bm \beta})^{\text T} \; ,
 \label{eq:posterior2}
\end{align}
where ${\bm K}({\bm \beta},{\bm \beta})$ is a $(M + 1) \times  (M + 1) $ matrix relating the covariance of observation points to each other, 
%. The ${\bm K}({\bm s}_*, {\bm s}) \in \mathbb{R}^{N_M \times N_N}$ term is the covariance matrix relating the test/prediction grid points to the observation points. Finally, the random vectors ${\bm f}_*$ and ${\bm f}$ represent the prediction/grid points and the observed values respectively. Note that ${\bm K}({\bm s}_*, {\bm s}) = {\bm K}({\bm s}, {\bm s}_*)^{\text T}$. ${\bm K}({\bm s}_*, {\bm s}_*) \in \mathbb{R}^{N_M \times N_M}$ is the covariance matrix relating the test/prediction grid points to each other.
\begin{equation}
{\bm K}({\bm s},{\bm s}) = 
\left[
\begin{array}{ccccc}
\kappa(\beta_{1},\beta_{1}) & \cdots & \kappa(\beta_{1},\beta_{M}) & 1 \\
\vdots & \ddots & \vdots & \vdots \\
\kappa(\beta_{M},\beta_{1}) & \cdots & \kappa(\beta_{M},\beta_{M}) & 1 \\
1 & \cdots & 1 & 0
\end{array}
\right] \;,
\label{eq:Kss}
\end{equation}
${\bm K}({\bm g},{\bm g})$ is a $G \times G$ matrix relating the covariance of grid points $\bm{g}$ to each other, i.e., the $i$th row and $j$th column is given by $[{\bm K}({\bm g},{\bm g})]_{ij} = \kappa(g_{i},g_{j})$ from \eqref{eq:kernel}, and ${\bm K}({\bm g},{\bm \beta})$ is the $G \times M$ matrix relating the covariance of grid points to samples. 
%Note that since the spatial positions of the measurements are shifted uniformly and the kernel depends only on the spatial lag between two measurements, the kernel is invariant to shifting the position of the measurements, that is, $\kappa(s_i,s_j) =\kappa(\beta_i(t),\beta_j(t))$. Hence,  the matrix \eqref{eq:Kss} and its inverse needs only to be computed once.
The data matrix $ {\bm Y}$  is included as a parameter in the expressions \eqref{eq:estimate}--\eqref{eq:posterior2} to emphasize that the spatial locations ${\bm \beta}$ and observations ${\bm f}$ are derived from $ {\bm Y}$ using \eqref{eq:msmt_correction}.

The GP estimate \eqref{eq:estimate} is computed in the wind-frame using available data just before takeoff (i.e., up to time $t_{N}$) at the grid points ${\bm g}$. The estimate in the $\mathcal{P}$ frame is converted into a time-varying estimate in the $\mathcal{I}$ frame via the relationship \eqref{eqn:wind}. That is, the wind-frame grid points ${\bm g}$ are shifted to inertial-frame points ${\bm g}+c t$
\begin{equation}
\label{eq:estimae_over_grid}
    \hat{\delta}_{{\rm N}}^\mathcal{I}(t,{\bm g}+c t; {\bm Y}) = \hat{\delta}_{{\rm N}}^\mathcal{P}({\bm g}; {\bm Y}) \;,
\end{equation}
where the $\hat{(\cdot)}$ indicates an estimate, the term $ct$ is added element-wise to the vector ${\bm g}$.
%and the notation from \eqref{eqn:wind} is slightly modified to emphasize the estimate is computed at grid points ${\bm g}$ using data ${\bm Y}$.
The grid points ${\bm g}$ are uniformly spaced over the interval $[0, \lambda]$, and linear interpolation is used to determine the estimated wind magnitude in-between grid points. 
%The total number of samples contained in the data \eqref{eq:msmtMatrix} is $N=MA$ and a high resolution of grid points (e.g., $G = 2000$ in this work) is used to estimate the wind-field. T
To account for the measurement noise \eqref{eq:msmt_noisy} the kernel function $\kappa(\beta, \beta')$ is modified by adding $\sigma_n^2$ to the expression in \eqref{eq:kernel} in cases where $\beta = \beta'$. The GP regression in this work makes the simplifying assumption that the location of each measurement is known precisely and the hyperparameters of the wind GP are known a priori. However, the approach can be extended to incorporate localization uncertainty \citep{jadaliha2012gaussian} and to learn hyperparameters in real-time \citep{rasmussen}.
% When estimating the wind at time $t$ the sample locations are shifted according to \eqref{eq:msmt_correction}. The estimate is then
% \begin{equation}\label{eq:estimate}
%     \hat{\delta}_{{\rm N}}(t,p_{{\rm N}}) = \begin{bmatrix} \bm{y}_{1:N}^{\text{T}} & 0 \end{bmatrix} {\bm{K}}(\overline{\bm{s}}^{k}_{i},\overline{\bm{s}}^{k}_{i})^{-1}  \bm{K}({\bm g},\overline{\bm{s}}^{k}_{i}) 
% \end{equation}
% or a specified inertial position $p_{{\rm N}}$ at time $t$.  For continuous time $t$, the estimation only uses the measurements \eqref{eq:msmtMatrix} available at the most recent time-step $t_k \leq t$.  At times $t_k \geq t$, the final aggregated set of measurements $\bm{y}_{1}, \cdots, \bm{y}_{N}$ are used within the estimation. 

To reduce computational complexity during GP regression a spatiotemporal truncation strategy may be employed \citep{Xu2011c}. Let $\varphi(t) = [- ct -\eta L, c(t_m-t)+ d +\eta L] $ denote a time-dependent spatial zone of acceptance in the wind-frame where measurements are included for regression with $\eta > 0$ being a chosen parameter. Measurements separated by three or more length scales have little influence on the estimates and thus $\eta \geq 3$ or greater is suggested. Only measurements associated with locations $\beta_i \in \varphi(t_{N})$ are used for GP regression. %described in Sec.~\ref{sec:msmt_correction}.

%(i.e., no GPS uncertainty). In principle, localization uncertainty can be incorporated as described in \citep{jadaliha2012gaussian}. It is also assumed that the hyperparameters of the wind GP are known. If these are unknown, they can be learned on-line as described in \citep{rasmussen}. The wind-field estimation is done in a centralized manner, which assumes that all quadrotors communicate their onboard measurements to a ground station.

%\subsection{Propagating Wind Measurements and Estimation}
%\label{sec:msmt_correction}
%Each anemometer collects and shares wind data that is aggregated at a centralized processing node that the quadrotor accesses, as detailed in eq. \eqref{eq:msmtMatrix}. Given that the anemometers are stationary in the inertial frame, it is assumed that their locations are known exactly throughout the simulation.  Each individual measurement can be defined for use in the GP regression by the set $\{t_i, s_i, y_i\}$, where $t_i$ is the time when the measurement took place, $s_i$ is the position of the anemometer in the inertial frame, and $y_i$ is the measured wind speed at point $s_i$.\\

% Depending on the initial location of the quadrotor $(p_{{\rm N},0})$, and the location of the closest anemometer in the operating region $(s_1)$, the quadrotor is set to take-off when the first measurement would reach its location, as defined by $(p_{{\rm N},0} - s_1)/c$, assuming that the quadrotor is down-stream from anemometer number one.

\subsection{Wind Regression Example}
To illustrate the GP regression with the convencted wind-field model, an example GP wind profile was generated in frame $\mathcal{P}$ using a mean $\mu_w = 8$~m/s, a GP length scale $L = 1.5$~m, and a GP variance $\sigma^2 = 4$~m$^2$.  The generated wind-field was then propagated through the simulated environment at a speed $c$, as described in Sec.~\ref{sec:wind_model}. Noisy measurements were taken by three stationary anemometers at a sampling frequency of 10 Hz with measurement noise $\sigma_n^2 = 0.6$~(m/s)$^2$. Once the first measurement reaches the initial position of the quadrotor, the trajectory planner uses the data collected for planning. The anemometers in the environment are denoted by the different colored square markers, and their respective measurements are shown by the same-colored star markers that propagate with the wind over time. Samples were taken from $0$ seconds to $t_{N} = 5$ seconds.  Three snapshots at 2, 8, and 14 seconds into the simulation are shown.
%show an improving estimate of the wind-field over time. 
%Given a takeoff time at 5 seconds, the estimate ceases to update, and continues to propagate with the wind-field for times $t > 5$ sec. 
\begin{figure}[H]
    \centering
    \includegraphics[width=\textwidth]{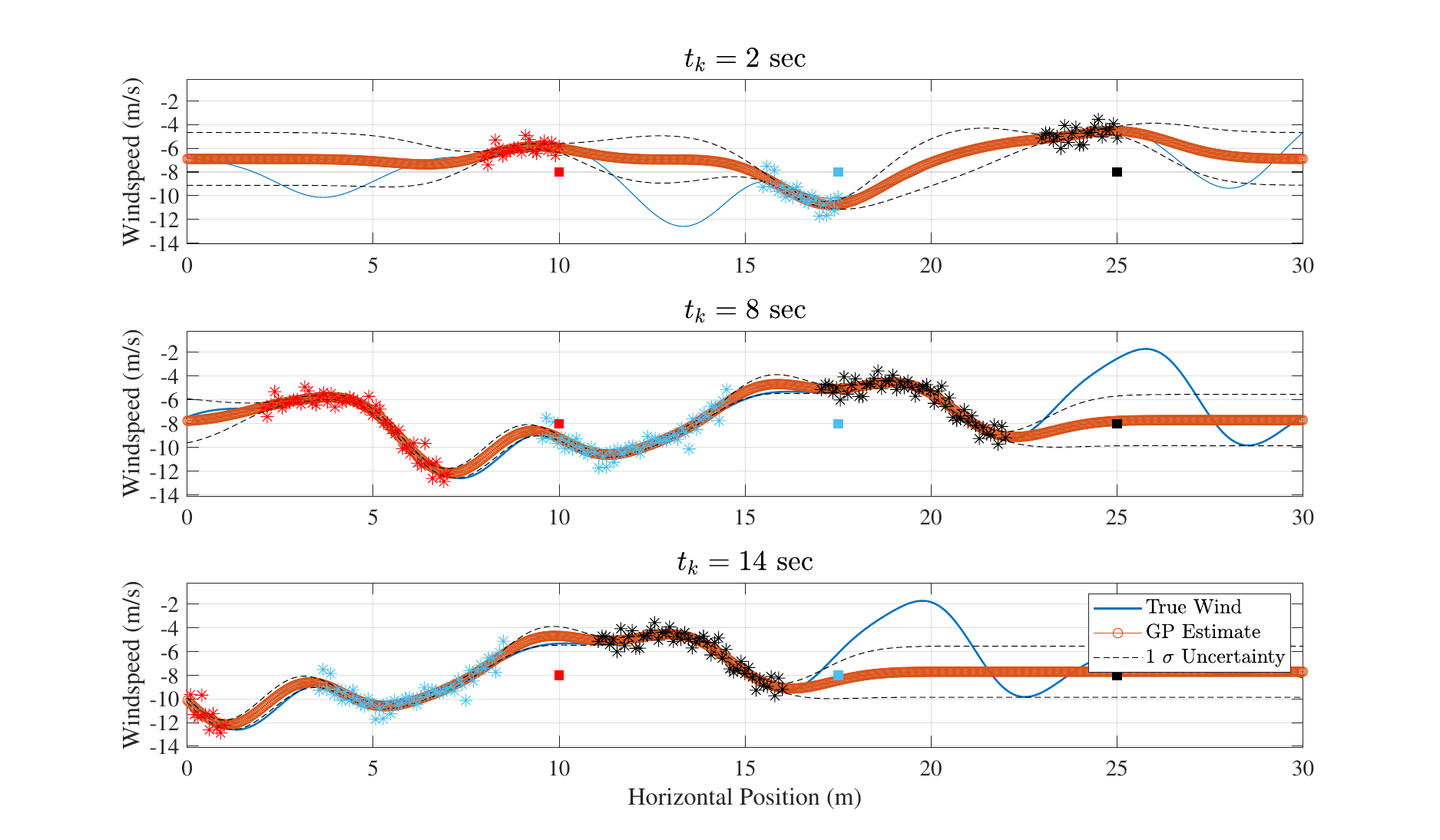}
    \caption{Example of the wind-field estimate evolving over the operating environment for three snapshots in time, with GP length scale $L = 1.5$ m, and GP variance $\sigma = 4 \text{ (m/s)}^2$.  The red, blue, and black squares represent the anemometers in the environment, and their respective measurements are shown as stars in the same colors (at a sampling frequency of 10 Hz with measurement noise $\sigma_n^2 = 0.6$~(m/s)$^2$).  The final estimate is made at time $t_N = 5$ sec.  and for times $t > t_N$ the estimated wind-profile is convected downstream. }
    %the estimate is t_{\rm{init}} = 5$ seconds and the estimated profile is 
    %convected downstream with speed $c$ for times $ t > 5$ sec.}
    %, as indicated by the estimate in the second and third plots no longer updating the estimate according to the measurements of the true wind-field.}
    \label{fig:exampleWindFig}
\end{figure}

\subsection{Proposed Estimation and Trajectory Planning Framework}\label{sec:framework}
\begin{figure}[H]
    \centering
    \includegraphics[width=0.9\textwidth]{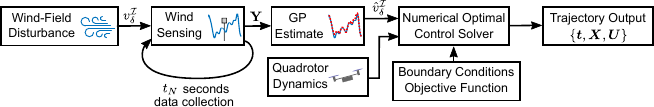}
    \caption{Proposed framework for leveraging wind-sensing infrastructure in an optimal control solver to plan a time-optimal trajectory to reach a desired waypoint.}
    \label{fig:framework}
\end{figure}

To address the time-optimal control problem of Sec. \ref{sec:probStatement}, the estimation and trajectory planning framework  sketched in Fig.~\ref{fig:framework} is proposed.  

A quadrotor is initialized in an environment with a wind-field disturbance, and with $A$ anemometers representing nearby wind sensing infrastructure that is available to the quadrotor.  The anemometers sampled the wind-field for $t_{N}$ seconds, as detailed in Sec.~\ref{sec:commModel}.  These measurements are used to generate a wind-field estimate, using the approach of Sec.~\ref{sec:gp} which is provided to an optimal control solver.  The modified quadrotor dynamics are
\begin{equation}
    %\dot{x} = 
    \begin{bmatrix}
        \dot{p}_{{\rm N}} \\ \dot{p}_{{\rm D}} \\ \dot{\theta} \\ \dot{u}_{{\rm r}} \\ \dot{w}_{{\rm r}}  \\ \dot{q}
    \end{bmatrix} = \begin{bmatrix}
        u_{{\rm r}} \cos{\theta} + w_{{\rm r}} \sin{\theta} + \hat{\delta}_{{\rm N}}(t,p_{{\rm N}}) \\ -u_{{\rm r}}\sin{\theta}+w_{{\rm r}}\cos{\theta} \\ q \\ -qw_{{\rm r}} \\ qu_{{\rm r}} \\ 0
    \end{bmatrix} + \begin{bmatrix}
        0 \\ 0  \\ 0 \\ f_1/m \\ {f_3}/m \\ \tau_2/I_{22}
    \end{bmatrix}
    \label{eq:simpEOM2}
\end{equation}
where $\hat{\delta}_{{\rm N}}(t_{k},p_{{\rm N},{k}})$ is the wind-field estimate sampled at the current position of the quadrotor $p_{{\rm N},{k}}$ (using linear interpolation over that grid points in \eqref{eq:estimae_over_grid}).  The dynamics \eqref{eq:simpEOM2} model the wind-field as a know time-varying disturbance for the purposes of trajectory planning.  The trajectory planning uses the numerical optimal control solver GPOPS-II \cite{gpops}, as described next.

\section{Simulation Results}\label{sec:results}
This section discusses the setup of several illustrative examples used to demonstrate the approach, implementation of the proposed framework of Sec. \ref{sec:framework} in the numerical optimal control solver GPOPS-II \cite{gpops},  and discusses  the results of the simulation trials comparing performance under different wind conditions and sensing characteristics.
%that  demonstrate the approach.
%For these simulations, the optimal control solver, GPOPS-II \cite{gpops} is used, for its versatility and ease of use with minimal tuning required.  
%Descriptions of the four simulation test cases cases are provided, and the implementation details for GPOPS-II are discussed.  Lastly, the Simulation results are presented and discussed.

\subsection{Simulation Setup}
\label{sec:control}
To evaluate the framework proposed in Sec. \ref{sec:framework} using GPOPS-II in MATLAB, six simulations were conducted that varied the wind mean $(\mu_w)$, GP variance $(\sigma^{2})$, GP length scale $(L)$, anemometer measurement noise $(\sigma_{n}^{2})$, and anemometer measurement frequency $(F)$, as shown in Table~\ref{tab:trials}. For each set of simulation parameters, a GP wind-field was generated by specifying $G$ uniformly spaced grid points $\bm{g} \in [0, \lambda]$ and hyperparameters ${\bm \theta} = [L, \sigma]^{\text T}$.  
%The maximum mission duration was defined as $t_m > t_{N}$.
%, where $t_{N}$ is the total time that the anemometers were set to sample the wind-field.
%, limiting the total measurements taken over time.  

\begin{table}[H]
    \footnotesize
    \centering
    \caption{Table of simulation trials used to evaluate the proposed estimation and trajectory planning framework.}
    \begin{tabular}{|c|c|c|c|c|c|l|}
        \hline
        \textbf{Trial} & \textbf{Wind mean, $\mu_w$} & \textbf{GP var., $\sigma^{2}$} & \textbf{GP scale, $L$} & \textbf{Sensor noise, $\sigma_{n}^{2}$} & \textbf{Sensor freq., $F$} & {\bf Wind Intensity / Sensor Quality} \\
        \hline
        \text{1} & $4 \text{ m/s}$ & $(\mu_{w}/4) \text{ m}^2$  & $(|c|t_{m}/10) \text{ m}$ & $0.6 \text{ (m/s)}^2$ & $10 \text{ Hz}$ & Low wind / higher-quality sensor \\
        \text{2} & $4 \text{ m/s}$ & $(\mu_{w}/4) \text{ m}^2$  & $(|c|t_{m}/10) \text{ m}$ & $1.2 \text{ (m/s)}^2$ & $2 \text{ Hz}$  & Low wind / lower-quality sensor \\
        \text{3} & $8 \text{ m/s}$ & $(\mu_{w}/2) \text{ m}^2$  & $(|c|t_{m}/20) \text{ m}$ & $0.6 \text{ (m/s)}^2$ & $10 \text{ Hz}$  & Moderate wind / higher-quality sensor \\
        \text{4} & $8 \text{ m/s}$ & $(\mu_{w}/2) \text{ m}^2$  & $(|c|t_{m}/20) \text{ m}$ & $1.2 \text{ (m/s)}^2$ & $2 \text{ Hz}$  & Moderate wind / lower-quality sensor \\
        \text{5} & $12 \text{ m/s}$ & $(\mu_{w}/2) \text{ m}^2$  & $(|c|t_{m}/20) \text{ m}$ & $0.6 \text{ (m/s)}^2$ & $10 \text{ Hz}$  & High wind / higher-quality sensor\\
        \text{6} & $12 \text{ m/s}$ & $(\mu_{w}/2) \text{ m}^2$  & $(|c|t_{m}/20) \text{ m}$ & $1.2 \text{ (m/s)}^2$ & $2 \text{ Hz}$  & High wind / lower-quality sensor \\
        \hline
    \end{tabular}
    \label{tab:trials}
\end{table}

Three anemometers were initialized at equal distances in the operating region, upstream of the initial location of the quadrotor.  
At the start of the simulation $(t = 0)$, the anemometers begin sampling the wind-field. 
%and the GP estimator generates an estimate at every sampling time.  
The first measurement reaches the quadrotor's initial position after $t_N = t_{\rm init} = (z_1 - p_{{\rm N},{\rm init}})/|c|$ seconds.
%where $z_1$ is the closest upstream anemometer and $p_{{\rm N},0}$ is the initial position of the quadrotor. 
At this time a GP estimate is computed and a trajectory is planned. 
%The quadrotor takes-off only when the first measurement reaches its initial position, which is defined as $(z_1 - p_{{\rm N},0})/c$ seconds, where $z_1$ is the closest upstream anemometer, and $p_{{\rm N},0}$ is the initial position of the quadrotor.  For simulation purposes, the GP estimates for the entire duration which the anemometers sample is pre-computed, and accessed in a look-up table during the GPOPS-II dynamics calculations. 
%At each time instant during the optimization, the wind-field estimate $\hat{\delta}_{{\rm N}}(t,p_{{\rm N},{k}})$ is sampled using linear interpolation based on the position $(p_{{\rm N}})$ of the quadrotor at time $t$.  

\subsection{Implementation in GPOPS-II}\label{sec:gpopsSimSetup}
GPOPS-II is an {hp}-adaptive version of the {Legendre-Gauss-Radau} (LGR) {orthogonal collocation method}, which uses Gaussian quadrature implicit integration with collocation performed at LGR points.  This process involves defining the upper and lower limits of the state, time, and control of the optimal control problem, the objective function, and the dynamics, then meshing a solution to the objective between the initial and final conditions using LGR orthogonal collocation.  Detailed information on the setup and use of GPOPS-II can be found in \cite{gpops,gpopsGuide}. Here we briefly discuss the general parameters used for the presented simulation results.  
%An in-depth description of how GPOPS-II functions can be found in \cite{gpops}.  
For this work, we use a mesh tolerance of $1\times10^{-7}$, and set the maximum number of iterations to 5. %, to decrease computational time.  
%Additionally, the wind-field is a time-varying disturbance, which adds to the complexity of attempting to optimize a solution for this problem.  
Simulations were conducted using the following boundary conditions:
\begin{equation}
\begin{aligned}
p_{{\rm N}}(t_{\rm{init}}) &= 5~{\rm m}  \\
p_{{\rm D}}(t_{\rm{init}}) &= 0~{\rm m} \\
\theta(t_{\rm{init}}) &= 0^\circ\\
u_{{\rm r}}(t_{\rm{init}}) &= 0~{\rm m/s} \\
w_{{\rm r}}(t_{\rm{init}}) &= 0~{\rm m/s}\\
q(t_{\rm{init}}) &= 0^\circ{\rm/s}
\end{aligned}
\qquad  \qquad \qquad 
\begin{aligned}
p_{{\rm N}}(t_{\rm{final}}) &= 15~{\rm m}  \\
p_{{\rm D}}(t_{\rm{final}}) &= -5~{\rm m} \\
-30^\circ \leq \theta(t_{\rm{final}}) &\leq 30^\circ \\
-5~{\rm m/s} \leq u_{{\rm r}}(t_{\rm{final}}) &\leq 5~{\rm m/s}\\
-5~{\rm m/s} \leq w_{{\rm r}}(t_{\rm{final}}) &\leq 5~{\rm m/s} \\
-100^\circ{\rm/s} \leq q(t_{\rm{final}}) &\leq 100^\circ{\rm/s}
\end{aligned} 
% \begin{aligned}
% p_{{\rm N}}(t_f) &= p_{{\rm N},f}  \\
% p_{{\rm D}}(t_f) &= p_{{\rm D},f} \\
% \theta(t_f) &= \theta_{\rm{min}} \leq \theta(t_f) \leq \theta_{\rm{max}} = -30^{\circ} \leq \theta(t_f) \leq 30^{\circ}\\
% u_{{\rm r}}(t_f) &= u_{{\rm r},\rm{min}} \leq u_{{\rm r}}(t_f) \leq u_{{\rm r},\rm{max}} = -5 \rm{ m/s} \leq u_{{\rm r}}(t_f) \leq 5 \rm{ m/s}\\
% w_{{\rm r}}(t_f) &= w_{{\rm r},\rm{min}} \leq w_{{\rm r}}(t_f) \leq w_{{\rm r},\rm{max}} = -5 \rm{ m/s} \leq w_{{\rm r}}(t_f) \leq 5 \rm{ m/s}\\
% q(t_f) &= q_{\rm{min}} \leq q(t_f) \leq q_{\rm{max}} = -100^{\circ} \rm{/s} \leq q(t_f) \leq 100^{\circ} \rm{/s}
% \end{aligned} \;,
\end{equation}
along with an initial time of $t_{\rm{init}} = 5$ sec. The final time was bounded as $t_{\rm{init}} \leq t_{\rm{final}} \leq 30$ sec., and the state was conservatively bounded according to 
\begin{align}
 -50~{\rm m} \leq p_{{\rm N}}(t) &\leq 50~{\rm m}  \\
 -50~{\rm m} \leq p_{{\rm D}}(t) &\leq 50~{\rm m} \\
-60^\circ \leq \theta(t) &\leq 60^\circ \\
-50~{\rm m/s} \leq u_{{\rm r}}(t) &\leq 50~{\rm m/s}\\
-50~{\rm m/s} \leq w_{{\rm r}}(t) &\leq 50~{\rm m/s} \\
-1000^\circ{\rm/s} \leq q(t) &\leq 1000^\circ{\rm/s}
\end{align}
for all $t \in (t_{\rm{init}}, t_{\rm{final}})$. The front/rear thrust values were bounded as $ 0 \leq T_{\rm f}, T_{\rm r} \leq T_{\rm max}$. All trials detailed in Table~\ref{tab:trials} were simulated for a single quadrotor with parameters listed in Table~\ref{tab:sim_params}.  Three anemometers were initialized to positions located at $z_{1,2,3} = \{10, 17.5, 25\}$~m.  GPOPS-II returns a planned trajectory $\{\bm{t},\bm{X},\bm{U}\}$, where $\bm{t}$ is a time vector spanning from the take-off time to the final time, $\bm{X}$ is a matrix containing the corresponding state history, and $\bm{U}$ is the control history used over the time span $\bm{t}$ to achieve $\bm{X}$. 

To verify the results of GPOPS-II solutions, the dynamics were re-simulated with explicit Runge-Kutta numerical integration (i.e., ODE45 in MATLAB \cite{ode45}), the known initial condition, the same estimated wind-field used in \eqref{eq:simpEOM2}, and spline interpolation across the control history $\bm{U}$ and times $\bm{t}$.  While in many cases, the output trajectory could be reproduced fairly accurately, we also encountered instances where significant discrepancies occurred.  These discrepancies may perhaps be attributed to interpolation or integration differences between ODE45 and GPOPS-II.
%solvers being explicit versus implicit, respectively. ODE45 uses a spline interpolated control based on the GPOPS-II provided trajectory.  
In this paper, only verified results are reported for which the ODE45 simulation closely matches the GPOPS-II trajectory (i.e., with the final state of both simulations reaching within 2\% of the desired waypoint when simulated with the same estimated wind conditions).  Once a solution was verified, the dynamics were again re-simulated in ODE45 using the true wind-field as the disturbance. The output of this latter simulation is referred to as the actual trajectory. 
%which were calculated based on the limited estimated knowledge of the wind-field.

%The quadrotor was initialized at $(p_{{\rm N},0},p_{{\rm D},0}) = (5~{\rm m},0~{\rm m})$ and provided a desired waypoint of $(p_{{\rm N},f},p_{{\rm D},f}) = (15,5)$, with the goal of flying from outside of the region where the anemometers are located, to within the region.    
\begin{table}[H]
    \footnotesize
    \centering
    \caption{Parameters used in simulations. The horizontal line separates vehicle dynamics parameters and wind-field model and estimation parameters.}
    \label{tab:sim_params}
    \begin{tabular}{|c|c|c|}
    \hline
        {\bf Parameter} & {\bf Symbol} & {\bf Value} \\
        \hline
        Quadrotor mass & $m$ & 3.696 kg \\
        Drag coefficients & ($C_{{\rm D},1}, C_{{\rm D},2}$)& (0.8, 0.4) \\
        Surface areas & $(A_1, A_3) $  & (0.0279, 0.109) ${\text m}^2$ \\
        Max. thrust bound & $T_{\text{max}}$ & 41.6964 N \\
        Air density & $\rho$ & 1.293 ${\text{kg}}/{\text{m}^3}$\\
        Inertia & $I_{22}$  & 0.0292 kg$\cdot \text{m}^2$\\
        Distance between propellers & $l$ & 0.254 m\\
        \hline
        Operating region length & $d$ & 40 m\\ 
        Number of grid points & $G$ & 2000 \\ 
        Truncation parameter & $\eta$  & 7\\
        Wind profile propagation speed & $c$  & $-$1 m/s\\
        Number of anemometers & $A$ & 3\\
        % Distance of Anem to Quad IC & $\delta$ & 5 m\\
        % Distance of Anem 1 to Anem 3 & $\gamma$ & 15 m \\
        % Wind Mean & $\mu_w$ & $\left[2, 4, 6\right] m/s$\\
        % GP variance scale (low/med/high) & $\sigma^{2}$ & $\left[\frac{\mu_w}{6}, \frac{\mu_w}{4}, \frac{\mu_w}{2}\right]$ \\
        % GP length scale (short/med/long) & $L$  & $\left[\frac{cT_{\text{sim}}}{20}, \frac{cT_{\text{sim}}}{10}, \frac{cT_{\text{sim}}}{5}\right]$ \\
        % Measurement noise & $\sigma_n^2$  & (0.2 m/s)$^{2}$\\ 
        % Measurement sampling time & $T_{\text sample}$ & 0.1 s \\
        \cline{1-3}
    \end{tabular}
\end{table}

\subsection{Results and Discussion}
The state and control history generated by GPOPS-II (i.e., the planned trajectory assuming the estimated wind-field) for each trial in Table~\ref{tab:trials} are presented alongside simulation of the control history with ODE45 using the true wind-field (i.e., the actual trajectory). 
%Each plot includes the trajectory generated by GPOPS-II and for
%For each plot, the trajectory was generated by GPOPS-II with knowledge of the wind estimate at times only prior to takeoff.  The control output from GPOPS-II was used to re-simulate the dynamics of the quadrotor \eqref{eq:simpEOM} using ODE45 while sampling from the true wind-field.  This process simulates the case of applying the proposed control framework to an actual system, and shows the possibility of real-world application.  Each set of plots includes a plot of the quadrotor trajectory using both GPOPS-II and ODE45, along with the state evolutions, and the control used for both solvers.  For each trial, 
The trajectory elapsed time $(t_{\rm final}-t_{\rm init})$ and the final Euclidean distance error between the planned and actual trajectories are reported.

\subsubsection{Trial 1: Low Wind Intensity, Higher-Quality Wind Sensor}
\begin{figure}[H]
    \centering
    \includegraphics[width=1\linewidth]{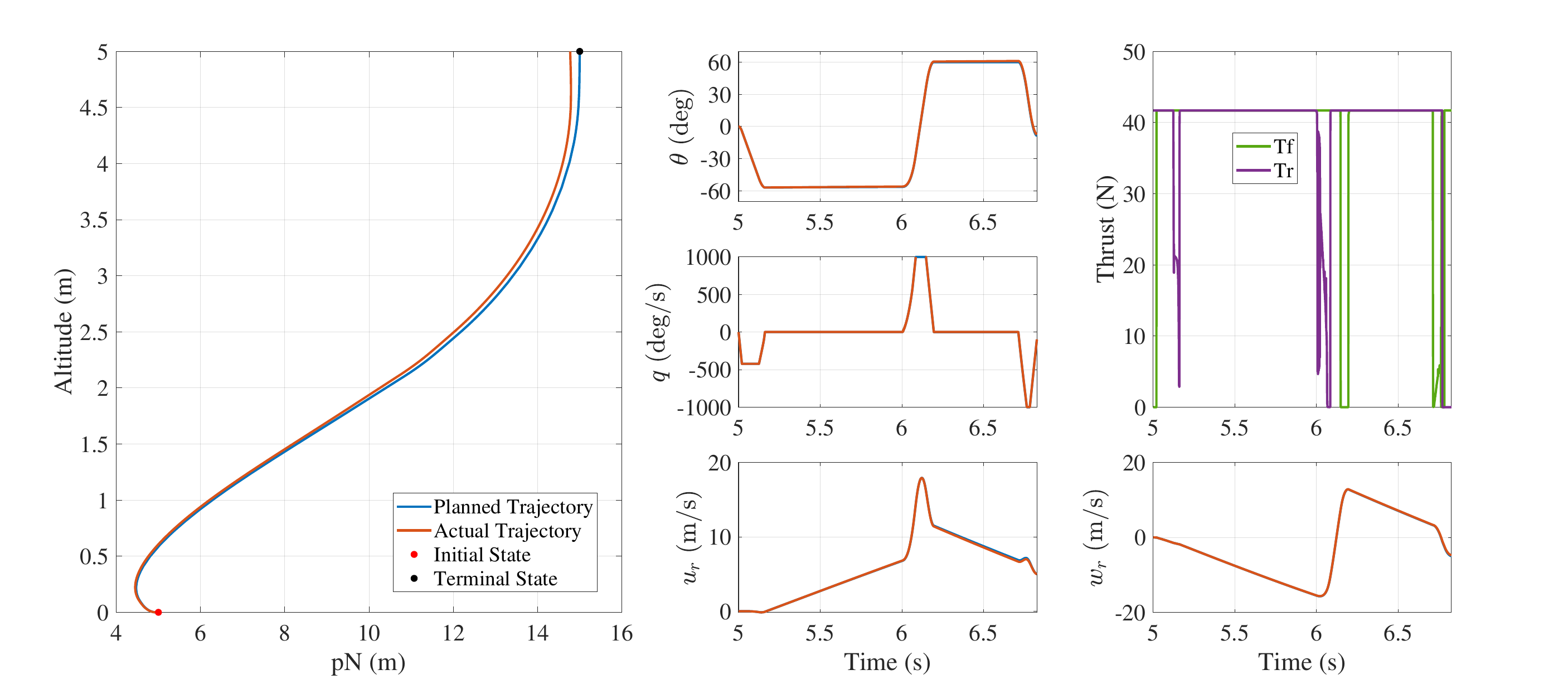}
    \caption{Simulation results for Trial 1. The left panel depicts the actual and planned trajectories in the vertical plane. The middle three, and bottom right panels show the pitch angle, pitch rate, and $u_{{\rm r}}, w_{{\rm r}}$ flow-relative velocities, respectively.  The upper right panel illustrates the control generated by GPOPS-II.}
    \label{fig:trial1}
\end{figure}

The results for Trial 1 are shown in Fig.~\ref{fig:trial1}. Since the quadrotor is initialized with zero velocity, pitch, and pitch rate, it is significantly perturbed when it encounters the wind disturbance on takeoff during the initial moments of the simulation. As can be seen in the plot of pitch angle over time the quadrotor initially pitches to its minimum allowable value ($-60^\circ$) --- holds this configuration while applying full thrust --- and then transitions to a maximum allowable pitch angle ($60^\circ$) just before levelling off as the terminal state is reached. The thrust controls vary rapidly during pitch change events, such as at 5.1 seconds, 6.1 seconds, and 6.8 seconds, with the pitch-rate saturating to the maximum allowable value ($\pm 1000^{\circ}$/s) during these times. This qualitative behavior of the controls and state history is similar across all trials. In this trial, with low wind and low sensor noise, the actual trajectory closely matches the planned trajectory and the final Euclidean distance error between them was $0.2259$~m. The planned trajectory cost was $(t_{\rm final} - t_{\rm init}) = 1.8254$ seconds.
%which as expected, with the lowest intensity wind-field, is the lowest of all the sets of trials.
%The goal of trial 1 was to evaluate the system with a moderately strong wind, which was sampled frequently with a moderately high measurement noise.  Due to the quadrotor being initialized with zero velocity, pitch, and pitch rate, it is greatly perturbed during the initial moments of the simulation.  This deviation is quickly fixed by the controls provided by GPOPS-II, through the use of a quick pitch in the direction of the waypoint, and the application of maximum thrust to get to the waypoint in minimal time.  The control efforts vary wildly during pitch change events, such as at 5.1 seconds, and 6.1 seconds.  The behaviour of the control is consistent throughout all trials, and is working as expected to change the pitch angle appropriately. The final euclidean distance error between the wind estimate aware GPOPS-II solution and the true wind ODE45 solution in this case was $0.2259\;m$.
% \begin{figure}[H]
%     \centering
%     \begin{subfigure}[b]{1\textwidth}
%         \includegraphics[width=1\linewidth]{figs/Trial1/trialNum1verificationRunRNG1.pdf}
%         \caption{Trial 1 - Verification Run}
%     \end{subfigure}
%     \begin{subfigure}[b]{1\textwidth}
%         \includegraphics[width=1\linewidth]{figs/Trial1/trialNum1performanceRunRNG1.pdf}
%         \caption{Trial 1 - Performance Run}
%     \end{subfigure}
%     \caption{Plots for trial 1.  For each set, the large leftmost plot depicts the trajectory taken by both solvers.  The middle three, and bottom right plots show the pitch angle, pitch rate, and $u_r, w_r$ flow-relative velocities, respectively.  The large plot on the right shows the control generated by GPOPS-II, and used in both solvers.}
%     \label{fig:trial1}
% \end{figure}

\subsubsection{Trial 2: Low Wind Intensity, Lower-Quality Wind Sensor}
\begin{figure}[H]
    \centering
    \includegraphics[width=1\linewidth]{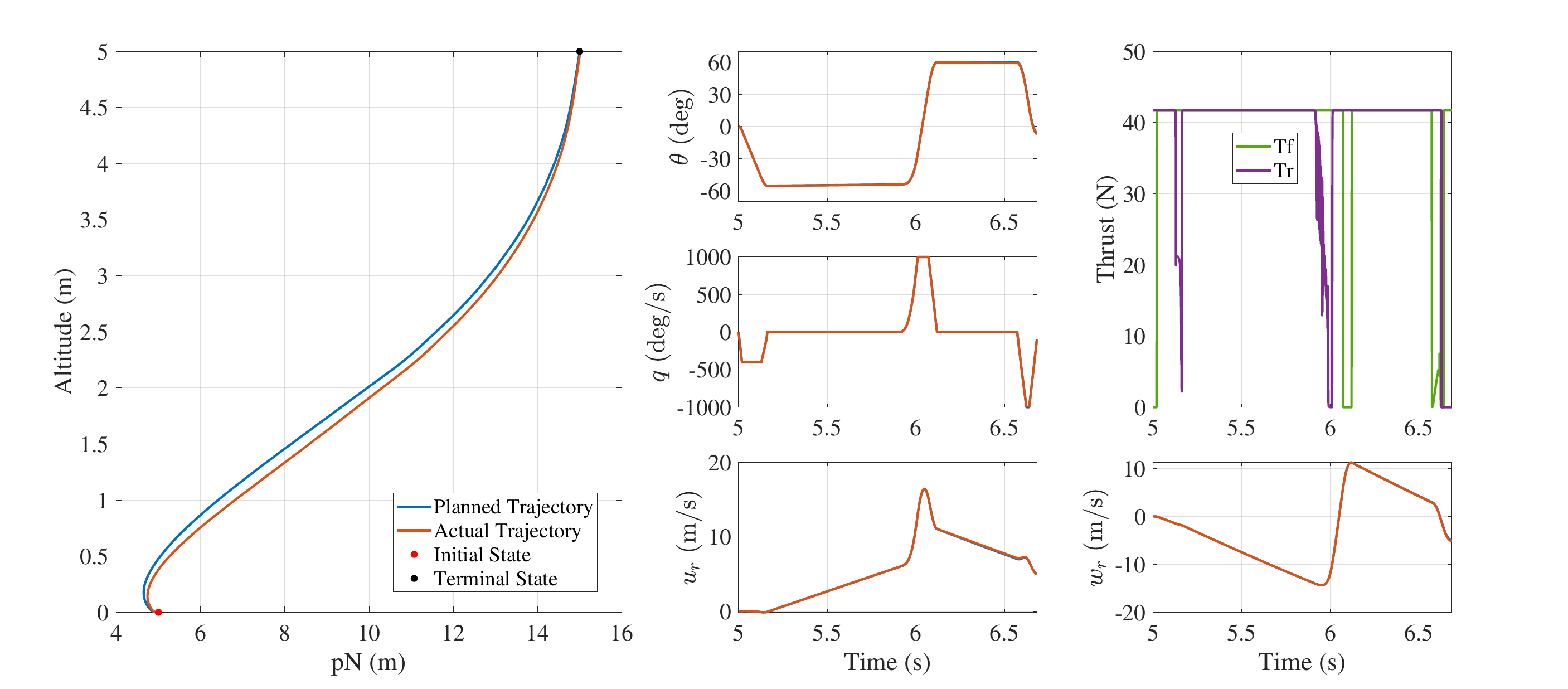}
    \caption{Simulation results for Trial 2. The left panel depicts the actual and planned trajectories in the vertical plane. The middle three, and bottom right panels show the pitch angle, pitch rate, and $u_{{\rm r}}, w_{{\rm r}}$ flow-relative velocities, respectively.  The upper right panel illustrates the control generated by GPOPS-II.}
    \label{fig:trial2}
\end{figure}

In Trial 2 (see Fig.~\ref{fig:trial2}) the measurement noise was increased, and the sampling frequency was decreased compared to Trial 1.  This trial simulates the effects of using a lower-quality wind sensor while holding the wind-field hyperparameters constant.  The overall trajectory is similar to Trial 1 and the final Euclidean distance error between the actual/planned trajectories at the final time is $0.0120$~m (less than but comparable to Trial 1). 
The planned trajectory cost was  $(t_{\rm final} - t_{\rm init}) = 1.6824$ seconds.
This result suggests that with a relatively low wind disturbance magnitude, the quadrotor was not adversely affected by a reduced measurement quality. The reduction in position error compared to Trial 1 is counter-intuitive, but it may be attributed to differences in numerical integration (especially over the rapidly changing control during pitch maneuvers). 
%The minimized-time for the trajectory calculated in trial 2 was 6.6824 seconds, which is slightly lower than the time calculated in trial 1.  This discrepancy between the calculated minimum-times between trials 1 and 2 can be attributed to the planner underestimating the time required  to reach the desired endpoint based on poor wind-field estimate information.
%since the performance run for this trial appears to be roughly the same as that of trial 1. The final euclidean distance error between the wind estimate aware GPOPS-II solution and the true wind ODE45 solution in this case was only $0.0120\;m$.  The difference in errors between trials 1 and 2 is unexpected, but may be attributed to numerical integration errors, as the two solvers perform numerical tasks differently. 
%Additionally, ODE45 uses a spline interpolated control based on the GPOPS-II provided trajectory, which may not apply exactly the same control efforts at the exact same time steps as the GPOPS-II solver.  Trials 1 and 2 performed roughly similar to each other, also indicating that this framework worked very well for wind-fields with low variance and mean.
% \begin{figure}[H]
%     \centering
%     \begin{subfigure}[b]{1\textwidth}
%         \includegraphics[width=1\linewidth]{figs/Trial2/trialNum2verificationRunRNG12.pdf}
%         \caption{Trial 2 - Verification Run}
%     \end{subfigure}
%     \begin{subfigure}[b]{1\textwidth}
%         \includegraphics[width=1\linewidth]{figs/Trial2/trialNum2performanceRunRNG12.pdf}
%         \caption{Trial 2 - Performance Run}
%     \end{subfigure}
%     \caption{Plots for trial 2.  For each set, the large leftmost plot depicts the trajectory taken by both solvers.  The middle three, and bottom right plots show the pitch angle, pitch rate, and $u_r, w_r$ flow-relative velocities, respectively.  The large plot on the right shows the control generated by GPOPS-II, and used in both solvers.}
%     \label{fig:trial2}
% \end{figure}

\subsubsection{Trial 3: Medium Wind Intensity, Higher-Quality Wind Sensor}
\begin{figure}[H]
    \centering
    \includegraphics[width=1\linewidth]{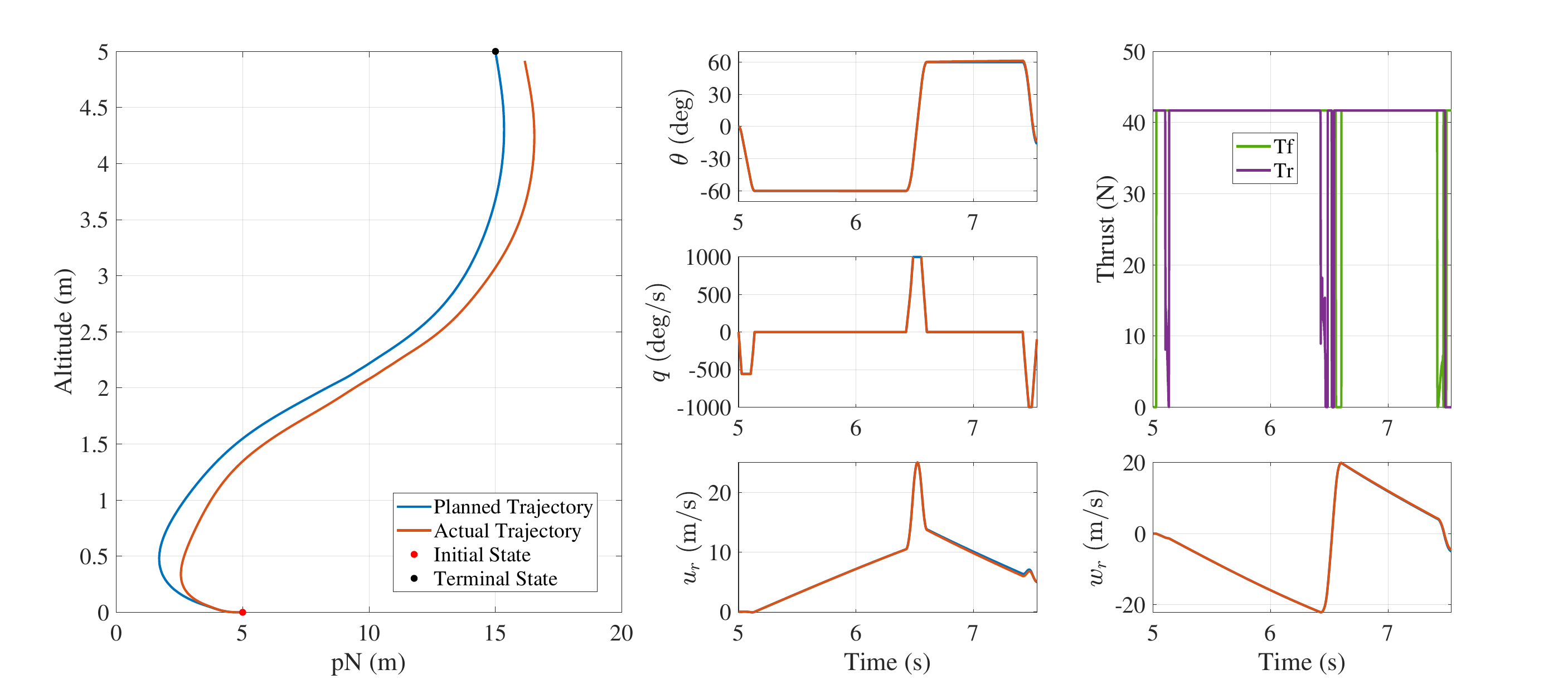}
    \caption{Simulation results for Trial 3. The left panel depicts the actual and planned trajectories in the vertical plane. The middle three, and bottom right panels show the pitch angle, pitch rate, and $u_{{\rm r}}, w_{{\rm r}}$ flow-relative velocities, respectively.  The upper right panel illustrates the control generated by GPOPS-II.}
    \label{fig:trial3}
\end{figure}

In Trial 3 (see Fig.~\ref{fig:trial3}) the wind mean was increased compared to Trials 1 and 2, with the same GP hyperparameters as used in Trial 1.  This change resulted in stronger wind  disturbance with more sudden gusts.  Fig. \ref{fig:exampleWindFig} shows a realization of the GP wind-field for Trials 3 and 4 using the same parameters.  The measurement characteristics used in Trial 3 were also the same as those in Trial 1. The final Euclidean distance error between the planned and actual trajectories was $1.1611$~m. The planned trajectory cost was $(t_{\rm final} - t_{\rm init}) = 2.5405$ seconds. In Trial 3, the actual trajectory is affected more greatly than that of Trial 1. For a wind-field with a greater mean, small errors in the estimated wind result amplified path deviations in the actual trajectory.  
%The minimized-time for the trajectory calculated in trial 3 was 7.5405 seconds, increasing from the time calculated for trials 1 and 2, as expected.
%With the higher mean, regardless of numerical integration differences between the two solvers, a greater error is expected between the two trajectories, as the stronger wind would accrue higher error over time with the estimate of the wind-field.  
% \begin{figure}[H]
%     \centering
%     \begin{subfigure}[b]{1\textwidth}
%         \includegraphics[width=1\linewidth]{figs/Trial3/trialNum3verificationRunRNG1.pdf}
%         \caption{Trial 3 - Verification Run}
%     \end{subfigure}
%     \begin{subfigure}[b]{1\textwidth}
%         \includegraphics[width=1\linewidth]{figs/Trial3/trialNum3performanceRunRNG1.pdf}
%         \caption{Trial 3 - Performance Run}
%     \end{subfigure}
%     \caption{Plots for trial 3.  For each set, the large leftmost plot depicts the trajectory taken by both solvers.  The middle three, and bottom right plots show the pitch angle, pitch rate, and $u_r, w_r$ flow-relative velocities, respectively.  The large plot on the right shows the control generated by GPOPS-II, and used in both solvers.}
%     \label{fig:trial3}
% \end{figure}

\subsubsection{Trial 4: Medium Wind Intensity, Lower-Quality Wind Sensor}
\begin{figure}[H]
    \centering
    \includegraphics[width=1\linewidth]{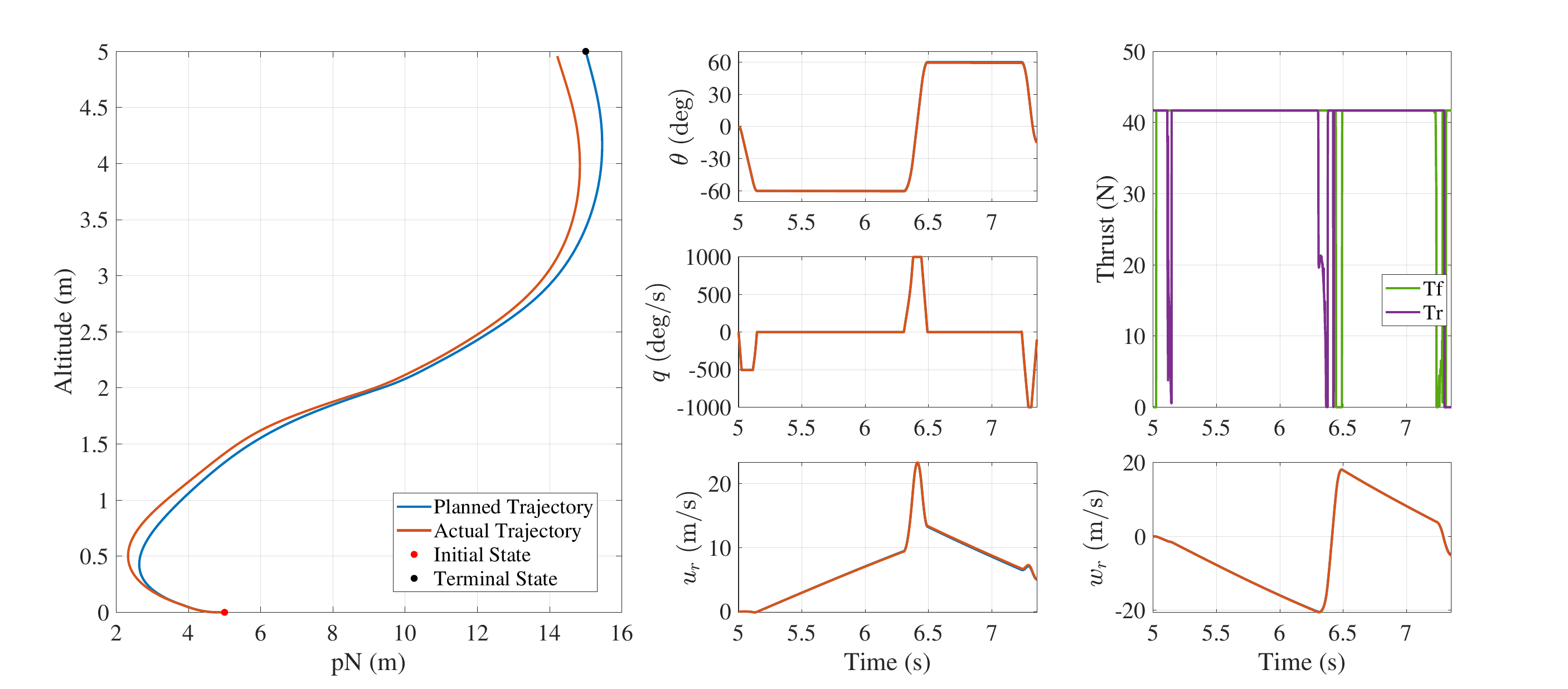}
    \caption{Simulation results for Trial 4. The left panel depicts the actual and planned trajectories in the vertical plane. The middle three, and bottom right panels show the pitch angle, pitch rate, and $u_{{\rm r}}, w_{{\rm r}}$ flow-relative velocities, respectively.  The upper right panel illustrates the control generated by GPOPS-II.}
    \label{fig:trial4}
\end{figure}

Trial 4 used the same moderate wind characteristics as Trial 3, but the lower-quality wind sensor.  The Euclidean position error for this trial ($0.7894$~m, see Fig.~\ref{fig:trial4}) is comparable to Trial 3. The planned trajectory cost was $(t_{\rm final} - t_{\rm init}) = 2.3542$ seconds. In comparison to the earlier Trials 1 and 2 at the lower mean wind setting, the results for Trials 3 and 4 in moderate wind approximately double the position error at the terminal state. Statistical analysis over a larger number of trials and refinement of the numerical integration strategy used in evaluating the actual versus planned trajectories would allow characterizing the effect of sensor quality in the low-to-moderate wind cases. 
%The minimized-time for the trajectory calculated in trial 4 was 7.3542 seconds, again decreasing between trials 3 and 4.

%used the same weak measurement characteristics as in trial 2.  This trial evaluated the effects of poor measurement on the estimate of a stronger, more varied wind-field.  The difference between the performance runs of trials 3 and 4 is immediately clear, in that the error appears to have doubled between the two trials.  This behaviour suggests that in a strong wind-field, with poor instrumentation, the proposed framework should not be utilized.   The final euclidean distance error between the wind estimate aware GPOPS-II solution and the true wind ODE45 solution in this case was $0.7894\;m$.  The error between cases 3 and 4 decreased slightly, but still may have decreased within the range of the numerical integration error between GPOPS-II and ODE45.  With this error difficult to quantify, it is difficult to state a specific reason.  

\subsubsection{Trial 5: High Wind Intensity, Higher-Quality Wind Sensor}
Similar to Trials 1 and 3, Trial 5 evaluates higher-quality sensing but with a stronger wind than previously simulated. The final Euclidean distance error was $1.0850$~m, larger than most of the previous cases, and still reasonably close the the desired waypoint. 
%The deviation between the planned and actual trajectory accrues rapidly during the initial pitch maneuver immediately after takeoff. 
The planned trajectory cost was $(t_{\rm final} - t_{\rm init}) = 3.105$ seconds.
%The minimized-time for the trajectory calculated in trial 5 was $8.105$ seconds, which is the largest of all of the trials.
%The stronger wind-field is causing ever larger errors between the trajectory generated with the estimated wind-field at takeoff, and the actual trajectory generated via perturbations by the true wind.  
%This framework has thus far provided good actual trajectory results within a reasonable range of the planned trajectory, but trial 5 shows that in strong winds, the framework may cause some large errors, even when frequent sampling is used to estimate the wind-field.
\begin{figure}[H]
    \centering
    \includegraphics[width=1\linewidth]{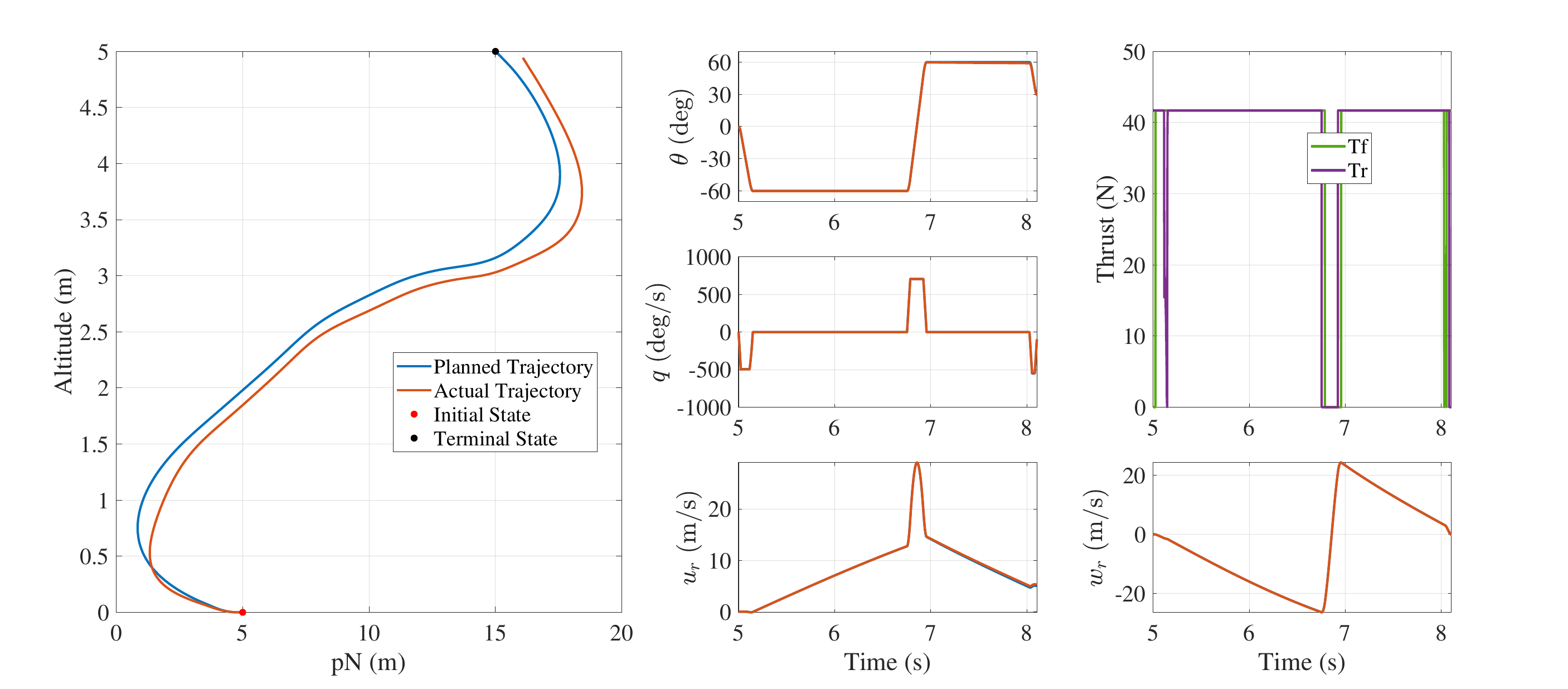}
    \caption{Simulation results for Trial 5. The left panel depicts the actual and planned trajectories in the vertical plane. The middle three, and bottom right panels show the pitch angle, pitch rate, and $u_{{\rm r}}, w_{{\rm r}}$ flow-relative velocities, respectively.  The upper right panel illustrates the control generated by GPOPS-II.}
    \label{fig:trial5}
\end{figure}

\subsubsection{Trial 6: High Wind Intensity, Lower-Quality Wind Sensor}
Lastly, Trial 6 evaluates a similar wind condition to Trial 5 but with a lower-quality wind sensor. The final Euclidean distance error between the planned and actual trajectories was $4.6731$~m --- the largest of all simulations presented and more than four times the error for Trial 5. For this strongest wind case, the wind sensor quality has a significant affect on position error. The planned trajectory cost was $(t_{\rm final} - t_{\rm init}) = 2.8504$ seconds. The lower cost of the trajectory for the case of a lower-quality wind sensor (compared to Trial 5) may be attributed to the decreased accuracy of the wind-field estimate. 
%The minimized-time for the trajectory calculated in trial 6 was $7.8504$ seconds, again following the trend of decreasing the calculated minimum-time due to a lower-quality estimate.
%This large error is what was expected for all prior trials with poor measurement characteristics.  Due to the strong wind-field, the error between the two trajectories accrued more rapidly, and is more readily apparent.  Trial 6 shows that the framework should not be used in strong wind-fields, when the wind sampling infrastructure can only provide poor quality measurements.
\begin{figure}[H]
    \centering
    \includegraphics[width=1\linewidth]{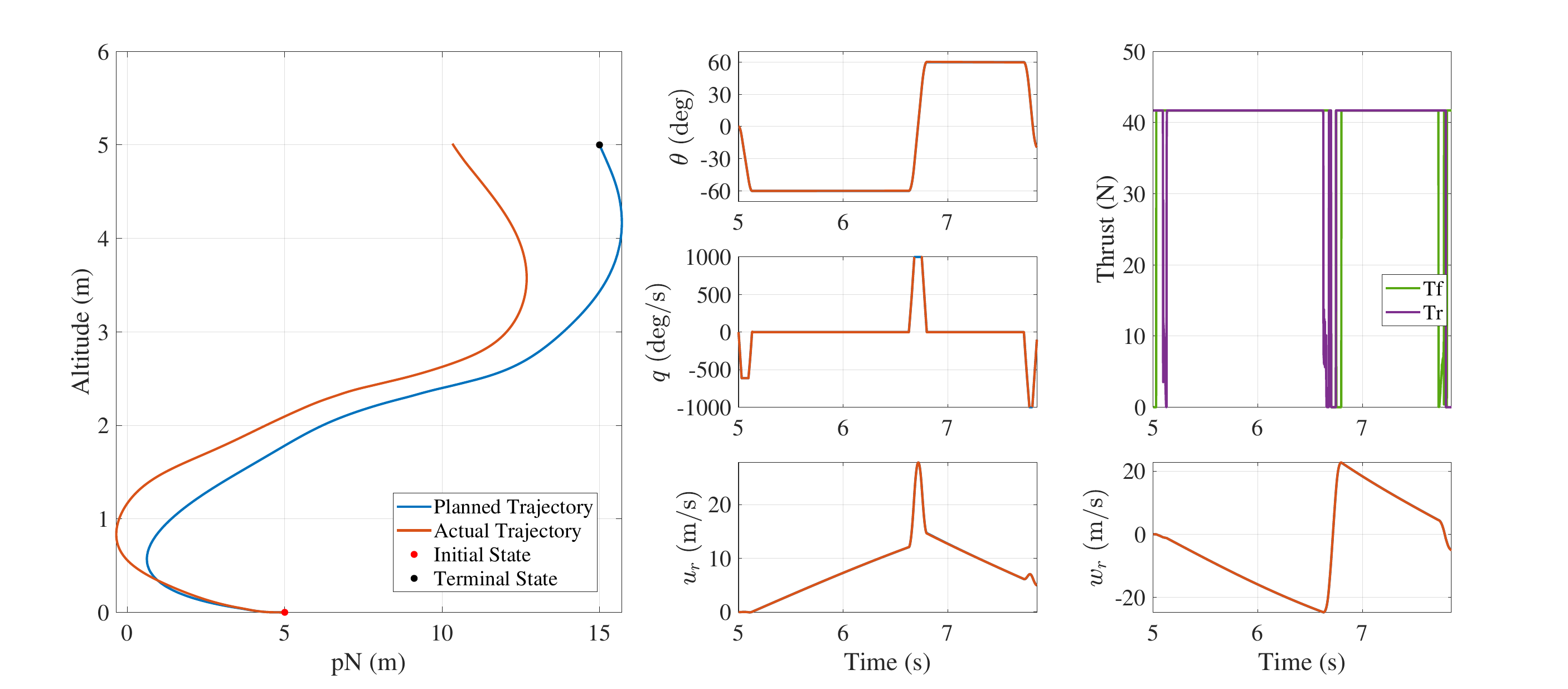}
    \caption{Simulation results for Trial 6. The left panel depicts the actual and planned trajectories in the vertical plane. The middle three, and bottom right panels show the pitch angle, pitch rate, and $u_{{\rm r}}, w_{{\rm r}}$ flow-relative velocities, respectively.  The upper right panel illustrates the control generated by GPOPS-II.}
    \label{fig:trial6}
\end{figure}
% \begin{figure}[H]
%     \centering
%     \begin{subfigure}[b]{1\textwidth}
%         \includegraphics[width=1\linewidth]{figs/Trial4/trialNum4verificationRunRNG1.pdf}
%         \caption{Trial 4 - Verification Run}
%     \end{subfigure}
%     \begin{subfigure}[b]{1\textwidth}
%         \includegraphics[width=1\linewidth]{figs/Trial4/trialNum4performanceRunRNG1.pdf}
%         \caption{Trial 4 - Performance Run}
%     \end{subfigure}
%     \caption{Plots for trial 4.  For each set, the large leftmost plot depicts the trajectory taken by both solvers.  The middle three, and bottom right plots show the pitch angle, pitch rate, and $u_r, w_r$ flow-relative velocities, respectively.  The large plot on the right shows the control generated by GPOPS-II, and used in both solvers.}
%     \label{fig:trial4}
% \end{figure}

\section{Conclusion and Future Work}\label{sec:conclusion}
An estimation and control framework was presented that supports trajectory planning in a one-dimensional uncertain wind-field.  The wind-field magnitude was modeled as a Gaussian Process (GP) that is spatially varying, with an unknown mean, and is convected downstream at a known constant speed.  Wind-sensing infrastructure in the operating area provides noisy measurements of the wind at upstream locations and is assimilated using Gaussian process regression to estimate the wind-field at unsampled locations and future time instants. The resulting GP estimate is used by a numerical optimal control solver (GPOPS-II) to compute a minimum-time trajectory to a desired vertical-plane  position.  The approach was evaluated over six trials that varied mean wind strength, wind strength variance, and wind-sensing measurement frequency and noise. The trials simulated the quadrotor following the optimized control in an open-loop fashion in the actual (true) wind field. Simulation results showed that the proposed approach is able to compensate for the wind-field in the operating environment and reach the desired waypoint reasonable well. The position error to the desired waypoint at the terminal time increased in the trials where the wind magnitude was larger ($8$ and $12$ m/s) compared to those where the wind magnitude was smaller ($4$ m/s). For the highest wind case, increasing measurement noise and reducing the sampling frequency led to lower quality wind estimate that decreased performance.
%as expected, but the quadrotor was still able to reach the vicinity of the desired waypoint. 
%For these simulations the commercial optimal control software GPOPS-II was used to calculate the optimal trajectory.  To fly the quadrotor in the true wind-field, the general differential equation solver ODE45 was used.  Each trial was validated by testing the generated control in ODE45 with the estimated wind-field, to ensure the two solvers agree under the same conditions.  Further, the performance of each trial was evaluated by calculating the euclidean distance error between the final position calculated by GPOPS-II and that of ODE45, when ODE45 used the true wind-field for its wind disturbances.  From these performance evaluations, in the trial which had a powerful and widely varying wind-field with poor measurement characteristics, the difference between the states of the solvers was the largest.  Alternatively, in a wind-field with the same properties but better measurement characteristics, the error was much smaller.  For the two trials with a more moderate wind-field, and both good and bad measurement characteristics, the performance errors were very small.  These results indicate that the proposed framework could feasibly be used in a real-world application, under the proper circumstances. 

Future work may consider more realistic wind-field models (e.g., extending to three-dimensions, consider an unknown convection speed) and other control objectives (e.g., trajectory tracking).  The approach could also be combined with wind-aware feedback-control strategies and multi-vehicle cooperative estimation and control. 

%Future work will include testing more wind and measurement cases in the presented simulation structure to further validate the results.  Additionally, the number of anemometers and their placements in the operation area can be varied, to determine the effect of larger gaps in the estimate.  The simulation may be changed from tracking a waypoint to tracking a path, to evaluate how the system performs over time, to evaluate both tracking performance and the effects of the optimal control solver essentially running out of valid wind estimates over time, due to the anemometers limited sampling times.  The wind-field can also be extended to two-dimensional effects, to evaluate the estimation framework in two dimensions.  Further different optimal control solver may be tested.  Nonlinear Model Predictive Control provides a good structure for live flight testing, and in-situ estimation, which could enable the estimation framework to change from stationary anemometers to anemometers mounted directly on the quadrotor.  Finally, work can be done on expanding the proposed simulations to include multiple quadrotor agents, to evaluate the ability of the framework to control multiple agents simultaneously in an uncertain wind-field.

\section*{Acknowledgments}
This work was  supported by NSF Grant No. 2301475.
%and startup funding provided by the University of North Carolina at Charlotte, The William States Lee College of Engineering, and the Department of Mechanical Engineering and Engineering Science.

\bibliography{refs.bib, refsAll_bibertool.bib}             % bib file to produce the bibliography

\end{document}